\documentclass[aps,preprint,nofootinbib,superscriptaddress]{revtex4}

\usepackage{graphicx}
\usepackage{amsmath}
\usepackage{amsfonts}
\usepackage{bm}
\usepackage{xcolor}
\usepackage{ulem}

\def\vec#1{{\bf #1}}

\newcommand{\bea}{\begin{eqnarray}}
\newcommand{\eea}{\end{eqnarray}}
\newcommand{\beq}{\begin{equation}}
\newcommand{\eeq}{\end{equation}}
\newcommand{\bqa}{\begin{eqnarray}}
\newcommand{\eqa}{\end{eqnarray}}

\newcommand{\nn}{\nonumber \\ }
\def\mqo2{{\!\!\!}}

\preprint{INT-PUB-11-026}
\begin{document}

\title{The three-boson system at next-to-leading order in
an effective field theory for systems with a large scattering length}
\author{Chen Ji}\email{jichen@phy.ohiou.edu}
\author{Daniel R. Phillips}\email{phillips@phy.ohiou.edu}
\affiliation{Department of Physics and Astronomy, Ohio University, Athens, OH\
45701, USA\\}
\author{Lucas Platter}\email{platter@chalmers.se}
\affiliation{Institute for Nuclear Theory, University of Washington, Seattle,
WA\ 98195, USA\\}
\affiliation{Fundamental Physics, Chalmers University of
  Technology, 412 96 G\"oteborg, Sweden}
\date{\today}

\begin{abstract}
  We analyze how corrections linear in the effective range, $r_0$,
  affect quantities in the three-body sector within an effective field
  theory for short-range interactions.  We demonstrate that
  observables can be obtained straightforwardly using a perturbative
  expansion in powers of $r_0$. In particular, we show that two
  linear-in-$r_0$ counterterms are needed for renormalization at this
  order if scattering-length-dependent observables are considered. We
  exemplify the implications of this result using various three-body
  observables. Analytic results for the running of the
  next-to-leading-order portion of the three-body force in this
  effective field theory are provided. Expressions which incorporate
  $O(r_0)$ corrections and relate the positions of features observed in
  three-atom recombination near a Feshbach resonance are presented.
\end{abstract}

\smallskip
\maketitle
\section{Introduction}
\label{sec:introduction}
Symmetries are one of the most important concepts in modern-day
physics. They are the foundation of the standard model and are also
frequently used as a basis for building new physical
theories. Even if a symmetry is only approximately fulfilled it can,
nonetheless, serve as a starting point for a systematic description of
physical observables, since the symmetry breaking effects may be
accounted for perturbatively. This approach has been successfully
employed in various effective field theories (EFTs) which are
low-energy expansions in the ratio of a small parameter over a large
parameter. The small parameter is frequently a small momentum and/or
energy associated with explicit symmetry breaking. A prominent example
of such an EFT is the chiral EFT, whose starting point is the chiral
limit of QCD. In this limit, pions are the Nambu-Goldstone bosons
associated with the spontaneously broken chiral symmetry of QCD. In the
chiral EFT the effect of the nonzero (up, down, and strange) quark
masses is then systematically included order-by-order in the
low-energy expansion.

While the use of symmetries to constrain theories is usually associated with
particle physics, or possibly many-body physics, it has also become 
important in other fields, such as non-relativistic few-body
physics. One example of an important symmetry in this area is discrete
scale invariance. Vitaly Efimov showed in 1970 that the non-relativistic
three-body system displays discrete scale invariance if the two-body
scattering length is large and the range of the interaction is
zero~\cite{Efimov70}. One well-known consequence is that in the limit
of infinite scattering length the ratio of the binding energies of two
successive three-body bound states is approximately 515.

Efimov also pointed out that his results apply to systems where
the two-body scattering length, $a$, obeys $|a| \gg \ell$, with $\ell$
the natural length-scale of the two-body
potential~\cite{Efimov70}. They are therefore relevant for a number of
systems. One example is nucleon-nucleon scattering, where the scattering length is large
compared to the range of the internucleon interaction. A second is provided by manipulation
of the atom-atom scattering length of trapped atoms by
an external magnetic field such that the atomic scattering displays a
Feshbach resonance. Few-body systems of nucleons and atoms, all with
interactions tuned such that a shallow two-body bound state is
present, will therefore display discrete scale invariance or the
remainders of this symmetry. For a recent review of this topic see,
e.g., Ref. \cite{Hammer:2010kp}.

Efimov's results have now been rederived as the leading-order (LO)
prediction of an EFT containing only short-range interactions~\cite{Bedaque:1998kg,Bedaque:1998km}. 
At LO this EFT corresponds to the Efimovian, large-$a$, $\ell=0$ limit.
We therefore refer to it hereafter as short-range EFT (SREFT). The contribution
of each subsequent order in the SREFT expansion is suppressed by an additional
power of $\ell/a$, with
the first corrections that
have to be included in this EFT being those resulting from a
finite two-body effective range, $r_0$. 
The effects of these
corrections have been analyzed over the last years in a number of
works \cite{Hammer:2001gh,Bedaque:2002yg,Platter:2006ev}. In these
studies effective-range corrections were considered for systems in
which the scattering length remains fixed. A full analysis requires,
however, that we allow for a variable scattering length. Such an
analysis was reported in \cite{Platter:2008cx,Ji:2010su}. In this
paper we lay out the derivation of the results reported in
\cite{Ji:2010su}. We discuss the renormalization of the SREFT in the
three-body sector at next-to-leading order (NLO) in the $\ell/a$
expansion. We show that if, and only if, scattering-length-dependent
observables are considered, an additional three-body counterterm is
required for renormalization at next-to-leading order. The analysis of
Ref.~\cite{Platter:2008cx} overlooked this result since the
combination of the non-perturbative treatment of effective-range
corrections and the cutoffs $\Lambda \gg 1/r_0$ employed there
modifies the ultraviolet properties of the theory.  The results of
Ref.~\cite{Platter:2008cx} therefore are strict EFT predictions only
in the limit $|r_0| \gg \ell$.

In Sec. \ref{sec:pionless-eft} we will introduce SREFT and briefly review what is known about the
two-body sector up to next-to-leading order. We then discuss, in
Sec.~\ref{sec:three-body-system}, the so-called modified
Skorniakov-Ter-Martirosian integral equation which constitutes the
application of the SREFT to the three-body sector at leading order.
Section~\ref{sec:three-body-observ} describes how NLO corrections
affect a variety of three-body observables.
Section~\ref{sec:subl-three-body} discusses the running of the NLO
parts of the three-body force, and presents analytic results for these
counterterms' renormalization-group evolution. Finally,
Sec.~\ref{sec:nlo-rel} presents relations between observables
measured in three-atom recombination which incorporate effects that
are NLO in SREFT. We end with a summary and outlook.

\section{The EFT for short-range interactions}
\label{sec:pionless-eft}
We employ the SREFT that describes non-relativistic particles
interacting through a finite-range interaction with a large scattering
length. The inverse of the range of the interaction sets the breakdown
scale for this EFT, which is constructed from contact interactions
alone. In nuclear physics this EFT is also known as the pionless
EFT (see Refs.~\cite{Beane:2000fx,Bedaque:2002mn} for reviews). However,
the SREFT can also describe other systems such as atoms close to a Feshbach
resonance. Since we perform our analysis
for identical bosons we will refer to the interacting particles as atoms.

At the heart of any EFT lies the Lagrangian. It includes all possible
interaction terms that are allowed by the symmetries of the underlying
interaction and is built from fields that correspond to the degrees of
freedom included in the EFT. The SREFT in its original form is
therefore built from atom fields alone, however, it has proven useful
to employ a Lagrangian that contains a dimer field. The relation
between these forms of the Lagrangian is elucidated in
Refs. \cite{Bedaque:1998km,Bedaque:1999vb,Braaten:2007nq,Braaten:2004rn}.
In these works the dimer field is sometimes chosen to be static,
sometimes dynamic.  The latter version is particularly convenient for
the inclusion of effective range corrections, and so we will work with
this form of the SREFT Lagrangian, written as:
\begin{equation}
\label{eq:Lagrangian}
\ensuremath{\mathcal{L}}=\psi^\dagger\left(i\partial_0 +
  \frac{\nabla^2}{2m}\right)\psi +\sigma T^\dagger\left(i\partial_0 +
  \frac{\nabla^2}{4m}-\Delta \right)T
-\frac{g}{\sqrt{2}}\left(T^\dagger
  \psi\psi+\textrm{h.c}\right)+hT^\dagger T \psi^\dagger \psi +
\ldots,
\end{equation}
where the ellipses represent additional higher-order interactions
which are suppressed at low momenta. As pointed out in
Ref.~\cite{Kaplan:1996nv}, a positive effective range can only be
described by this theory if $\sigma=-1$. In this case the free theory
describes an atom and a ghost field. We will employ the Lagrangian
(\ref{eq:Lagrangian}) and will expand all elements of all Feynman
graphs in powers of $r_0$. The SREFT
Lagrangian is then equivalent, order-by-order in the $r_0$ expansion, to
the SREFT Lagrangian with an atom field alone, as presented in, e.g.,
Refs.~\cite{Beane:2000fx,Bedaque:1999vb}.

\begin{figure}[tbpic:dimer]
\centerline{\includegraphics[width=15cm,angle=0,clip=true]{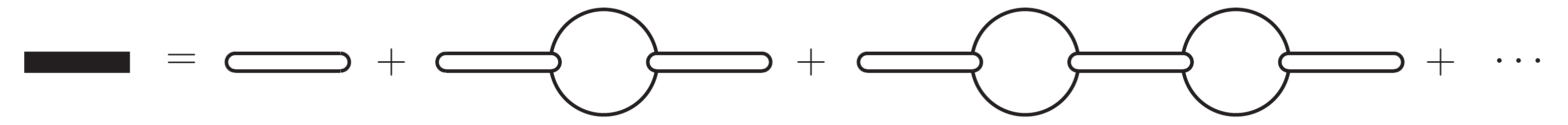}}
\caption{The series that is summed to obtain the dressed dimer propagator, leading to Eq.~(\ref{eq:prop}).}
\label{pic:dimer}
\end{figure}

The Feynman rules are derived from Eq.~\eqref{eq:Lagrangian} and the
atom propagator in momentum space is
\begin{equation}
iS(p_0,p)=\frac{i}{p_0-\frac{p^2}{2m}+i\epsilon},
\end{equation}
where $p_0$ is the energy and $p=|{\bf p}|$.
The large scattering length leads to large loop effects, 
and the EFT power counting requires therefore that the two-body
interaction is iterated to all orders (Fig.\ref{pic:dimer}).
The resulting dressed dimer propagator is
\begin{equation}
\label{eq:prop}
i\mathcal{D}(p_0,p)=\frac{-i}{p_0-\frac{p^2}{4m}-\Delta+\frac{mg^2}{4\pi}\sqrt{
-mp_0+\frac{p^2}{4}-i\epsilon}+i\epsilon}.
\end{equation}
If the scattering length $a$ is positive, the two-body system will display
a bound state ({\it shallow dimer}). Equation~\eqref{eq:prop}
therefore has to have a pole at the on-shell four-momentum
$p_0=\frac{p^2}{4m}-\frac{\gamma^2}{m}$, where $-\frac{\gamma^2}{m}$
is the binding energy of the shallow dimer. We can rewrite this
condition and obtain
\begin{equation}
 -\gamma^2 - m\Delta + \frac{m^2 g^2}{4\pi}\gamma=0,
\end{equation}
which has the solution
\begin{equation}
\label{eq:gamma-Delta}
\gamma=\frac{m^2g^2}{8\pi}\left(1-\sqrt{1-\frac{64\pi^2\Delta}{m^2g^2}}\right).
\end{equation}
We can relate the coupling constants $g$ and $\Delta$ to scattering
length $a$ and effective range $r_0$ by using the S-wave effective
range expansion
\begin{equation}
\label{eq:rel-a-gamma}
-\frac{1}{a}=-\gamma+\frac{1}{2}r_0\gamma^2
\end{equation}
which leads to $a=\frac{mg^2}{4\pi}\frac{1}{\Delta}$ and
$r_0=\frac{8\pi}{m^2g^2}$. Note that $\sigma=-1$ was used in
Eq.~\eqref{eq:Lagrangian} to derive these results.
In terms of these quantities, 
the wave-function
renormalization factor $Z$ for this bound state is:
\begin{equation}
\label{eq:Zfactor}
\frac{1}{Z}=i\frac{\partial}{\partial
p_0}(i\mathcal{D}(p))^{-1}|_{p_0=\frac{p^2}{4m}-\frac{\gamma^2}{m}}=
\frac{m^2g^2}{8\pi\gamma}(1-r_0\gamma).
\end{equation}

Thus, the dressed dimer propagator can be reexpressed as
\begin{equation}
\label{eq:2}
i\mathcal{D}(p_0,p)=\frac{-i4\pi/mg^2}{-\gamma+\frac{1}{2}
r_0(\gamma^2+mp_0-p^2/4)+\sqrt{-mp_0+p^2/4-i\epsilon}+i\epsilon}.
\end{equation}

Two-body scattering phaseshifts for relative momentum $k$ can be
obtained from the propagator $\mathcal{D}(p_0,p)$ by evaluating it at
the on-shell point of two atoms scattering with center-of-mass
momentum ${\bf p}=0$ and $E=k^2/m$, and multiplying it by $-ig^2$. This
yields a two-body amplitude:
\begin{equation}
\frac{4 \pi}{m} \frac{1}{\gamma - \frac{1}{2} r_0 (\gamma^2 + k^2) + i k},
\end{equation}
i.e. one in conformity with the effective-range expansion around the two-body
bound-state pole.

However, the propagator in Eq.~(\ref{eq:2}) cannot be directly
employed in an integral equation for three-body physics, since it
contains a spurious pole. This deep bound state (pole at momentum 
$\sim 1/r_0$) arises since we have chosen to work
with a ghost field that leads to unphysical features in the theory,
such as negative-norm states and/or violations of unitary. However, when
we expand the propagator in powers of $\gamma r_0$, the deep pole does
not appear at any finite order:
\begin{equation}
i\mathcal{D}(p_0,p)=\sum_n
\frac{-i4\pi/mg^2}{-\gamma+\sqrt{-mp_0+p^2-i\epsilon}+i\epsilon}
\left(\frac{r_0}{2}\right)^n \left(\gamma+\sqrt{-mp_0 +p^2/4}\right)^n.
\end{equation}
The leading-order (LO) dimer propagator is therefore given by
\begin{equation}
i\mathcal{D}^{(0)}(p_0,p)=\frac{-i4\pi/mg^2}{-\gamma+\sqrt{-mp_0+p^2-i\epsilon}
+i\epsilon},
\end{equation}
and the next-to-leading order $\ensuremath{\mathcal{O}}(\gamma r_0)$ correction
is
\begin{equation}
i\mathcal{D}^{(1)}(p_0,p)=-i\frac{4\pi}{mg^2} \times \frac{r_0}{2}
\frac{\gamma+\sqrt{-mp_0+p^2/4}}{-\gamma+\sqrt{-mp_0+p^2/4}}.
\end{equation}

\section{Three-Body Amplitudes}
\label{sec:three-body-system}
The calculation of three-body observables requires the solution of an
integral equation which we will call the modified
Skorniakov-Ter-Martirosian equation. It is the Faddeev equation for
non-relativistic particles interacting through two-body and three-body
zero-range interactions. We will show this equation below.  Details of
its derivation can be found in
Refs.~\cite{Bedaque:1998kg,Bedaque:1998km}.  A three-body interaction
is included there, to ensure cutoff-independent results. The
running of the associated coupling constant ($h$) displays a limit
cycle.

Our analysis concerns the corrections to the modified STM equation and is based on
rewriting all involved quantities in the form
\begin{eqnarray}
\label{eq:expansion}
\mathcal{D}(p)&=&\mathcal{D}^{(0)}(p)+\mathcal{D}^{(1)}(p)+\ldots\nn
t(q,p)&=&t^{(0)}(q,p)+t^{(1)}(q,p)+\ldots\nn
H(\Lambda)&=&H_0(\Lambda)+H_1(\gamma,\Lambda)+\ldots,
\end{eqnarray}
where $H(\Lambda)=\Lambda^2 h/2mg^2$, and each quantity in
Eq.~\eqref{eq:expansion} is expanded in powers of $k\, r_0$ and
$\gamma r_0$. Hence all quantities in the calculations that follow are
computed up to next-to-leading order in the $\ell/a$
expansion. Since $\gamma , k \sim 1/a$ and we assume $r_0 \sim \ell$,
this implies that our goal in this work is to compute $t^{(1)}$ and
associated NLO parts of physical observables in this expansion.  The
$t$-matrix obtained in this way is then related to the renormalized
amplitude for s-wave atom-dimer scattering at relative momentum $k$,
$T(k)$, via:
\begin{equation}
T(k)=\sqrt{Z}(t_0(k,k;E) + t_1(k,k;E) + \ldots) \sqrt{Z},
\label{eq:Tk}
\end{equation}
where, for elastic atom-dimer scattering, the magnitude of the
relative incoming momentum $|\vec{k}|$ equals the relative outgoing
momentum $|\vec{p}|$ and the three-body energy, $E$, is given by:
\begin{equation}
E=\frac{3k^2}{4m}-\frac{\gamma^2}{m}.
\label{eq:Eandk}
\end{equation}
In Eq.~(\ref{eq:Tk}), $Z$ is the wave-function
renormalization factor given in Eq.~\eqref{eq:Zfactor}. 
We also expand this $Z$-factor in powers of $\gamma r_0$:
\begin{equation}
Z_0=\frac{8\pi\gamma}{m^2g^2}~, \quad Z_1=\frac{8\pi\gamma^2 r_0}{m^2g^2},
\end{equation}
and so on, thereby generating an expansion for $T(k)$ analogous to those listed
in Eq.~(\ref{eq:expansion}).

\begin{figure}[tbpic:t1]
\centerline{\includegraphics[width=16cm,angle=0,clip=true]{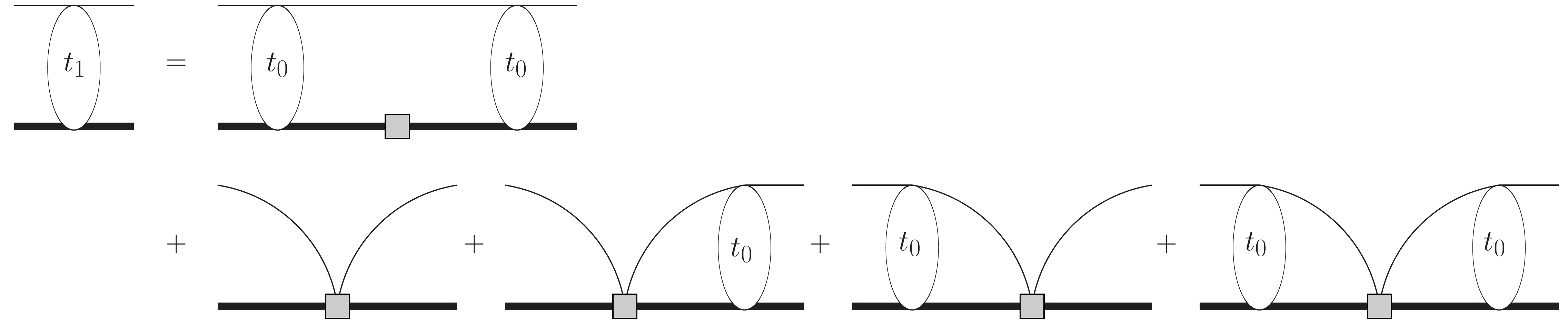}}
\caption{The diagrams for the next-to-leading order t-matrix.
  Propagators and vertices with gray squares denote 
  NLO corrections.}
\label{pic:t1}
\end{figure}

The leading-order three-body amplitude, $t_0$, is computed by iterating the
two- and three-body interactions of Eq.~(\ref{eq:Lagrangian}) to all
orders via the integral equation~ \cite{Bedaque:1998km}
\begin{equation}
\label{eq:t0}
\tilde{t_0}(q,p;E)=M(q,p;E)+\frac{2}{\pi}\int_0^\Lambda dq'\
\frac{q'^2}{-\gamma+\sqrt{3q'^3/4-mE-i\epsilon}} M(q',p;E) \tilde{t_0}(q,q';E),
\end{equation}
where
\begin{equation}
 M(q,p;E)=\frac{1}{qp} \log\left(
  \frac{q^2+p^2+qp-mE}{q^2+p^2-qp-mE}\right)
+\frac{2H_0(\Lambda)}{\Lambda^2}. 
\end{equation}
Note that we have rescaled the t-matrix in Eq.~\eqref{eq:t0}
$t_0(q,p;E) = mg^2 \tilde{t}_0(q,p;E)$ such that it depends only on
physical quantities.  The integral equation without the three-body
force was first derived by Skorniakov and Ter-Martirosian and
corresponds to three particles interacting through zero-range two-body
interactions~\cite{STM57}.

The LO three-body force, $H_0(\Lambda)$ is fixed by fitting it to
one three-body observable. A combination of analytical arguments and numerical
studies then show that $H_0(\Lambda)$ varies with $\Lambda$ as:
\begin{equation}
  \label{eq:1}
  H_0(\Lambda)=c \frac{\sin(s_0 \ln(\Lambda/\bar{\Lambda})+\arctan(s_0))}
{\sin(s_0\ln(\Lambda/\bar{\Lambda})-\arctan(s_0))},
\end{equation}
The constant $c$ is $\mathcal{O}(1)$ and depends on details of the regularization of the 
modified STM equation. For the regulator employed in Eq.~\eqref{eq:t0} we confirm the value
$c=0.879$ determined numerically in 
Ref.~\cite{Braaten:2011sz}.

We calculate corrections to the amplitude obtained from the LO modified STM
equation, $\tilde{t}_0$, by considering
diagrams with 
a single insertion of the NLO dimer propagator. It was shown by
Hammer and Mehen \cite{Hammer:2001gh} that such a perturbative
inclusion of effective range corrections also requires the insertion
of a subleading, energy-independent, three-body force. In
Fig.~\ref{pic:t1} we display the diagrams that have to be evaluated. The
application of the Feynman rules gives, for the first-order correction
to the amplitude 
\begin{eqnarray}
\label{eq:nlo-tmatrix1}
it^{(1)}(\vec{q},\vec{p};E)&=&\int \frac{d^4 q'}{(2\pi)^4}
iS(E-q'_0,q')i\mathcal{D}^{(1)}(q'_0,q')\nn &&\times
it^{(0)}(\vec{q},\vec{q'},q'_0 -E +\frac{q^2}{2m})\
it^{(0)}(\vec{q'},\vec{p},E-\frac{p^2}{2m}-q'_0)\nn
&+&i\frac{2mg^2 H_1(\gamma,\Lambda)}{\Lambda^2} \nn
&&\times \left[1+\int \frac{d^4q'}{(2\pi)^4}\ i\mathcal{D}^{(0)}(q'_0,q')\
iS(E-q'_0,q')\ it_0(\vec{q},\vec{q'},q'_0 -E+\frac{q^2}{2m})\right]\nn
&&\times \left[1+\int \frac{d^4q'}{(2\pi)^4}\ i\mathcal{D}^{(0)}(q'_0,q')\
iS(E-q'_0,q')\ it_0(\vec{q'},\vec{p},E-\frac{q^2}{2m}-q'_0)\right].\nn
\end{eqnarray}
The complete NLO correction to the S-wave projection of the t-matrix
is therefore
\begin{eqnarray}
\label{eq:t1}
\tilde{t_1}(q,p;E)
&=&\frac{1}{\pi}\int_0^\Lambda dq'
q'^2\frac{\gamma+\sqrt{3q'^2/4-mE}}{-\gamma+\sqrt{3q'^2/4-mE-i\varepsilon}}\tilde{t_0}(q,
q';E) \tilde{t_0}(q',p;E)\nn
&&+\frac{2 
\tilde{H_1}(\gamma,\Lambda)}{\Lambda^2}\left[1+\frac{2}{\pi}\int_0^\Lambda
dq'\frac{q'^2}{-\gamma+\sqrt{3q'^2/4-mE-i\varepsilon}}\tilde{t_0}(q,q';E)\right]\nn
&&\hspace{2.5cm}\times\left[1+\frac{2}{\pi}\int_0^\Lambda
dq'\frac{q'^2}{-\gamma+\sqrt{3q'^2/4-mE-i\varepsilon}}\tilde{t_0}(q',p;E)\right],
\end{eqnarray}
where $\tilde{t}_1$ and $\tilde{H}_1$ are defined via $t_1(q,p;E) \equiv r_0
mg^2 \tilde{t_1}(q,p;E)$, and $ H_1 \equiv r_0 \tilde{H_1}$.

The expression above is similar to the one obtained by Hammer and
Mehen in Ref.~\cite{Hammer:2001gh}. However, in their work one
contribution from the contour integration in
Eq.~\eqref{eq:nlo-tmatrix1} was erroneously omitted. In that
approximation the first term in Eq.~(\ref{eq:t1}) becomes:
\begin{equation}
\frac{1}{\pi}\int_0^\Lambda dq'
q'^2\frac{2 \gamma}{-\gamma+\sqrt{3q'^2/4-mE-i\varepsilon}}\tilde{t_0}(q,
q';E) \tilde{t_0}(q',p;E).
\end{equation}
As will become clear below, this omission leads to the conclusion that only one
counterterm is required to renormalize the three-body problem in SREFT
at NLO. While this works in practice for fixed-$a$ observables, such
as the neutron-deuteron phase shifts computed in
Ref.~\cite{Hammer:2001gh}, it produces incorrect results if
experiments in which $a$ varies, like those examining loss features in
gases of cold atoms, are analyzed.

$\tilde{t}_0$, $\tilde{t}_1$, etc. are 
generally complex when scattering states are considered, albeit real when
the bound-state problem is considered. It is therefore convenient to
introduce the real K-matrix, which contains the same information as the
t-matrix but is easier to calculate. The LO half-on-shell K-matrix
obeys an STM equation in which the $i\epsilon$ prescription is
replaced by a principal-value integration (indicated by $\mathcal{P}$)
\begin{equation}
\label{eq:K0-half}
\tilde{K_0}(k,p;E)=M(k,p;E) +\frac{8}{3\pi}
\ensuremath{\mathcal{P}}\int_0^\Lambda dq'\ \frac{q'^2
(\gamma+\sqrt{3q'^3/4-mE})}{q'^2-k^2} M(q',p;E) \tilde{K_0}(k,q';E).
\end{equation}
The half-on-shell t-matrix at LO is related to the K-matrix via
\begin{equation}
\label{eq:t0-K0-half}
\tilde{t_0}(k,p;E) = \frac{\tilde{K_0}(k,p;E)}{1-i\frac{8\gamma
k}{3}\tilde{K_0}(k,k;E)}.
\end{equation}
The fully-off-shell t-matrix at leading order is also related to the K-matrix
through a similar transformation
\begin{equation}
\label{eq:t0-K0}
\tilde{t_0}(q,p;E) = \tilde{K_0}(q,p;E) + i\frac{8\gamma k}{3}
\tilde{K_0}(k,p;E) \tilde{t_0}(q,k;E).
\end{equation}
This allows us to write the fully-off-shell K-matrix as
\begin{equation}
\label{eq:K0-full}
\tilde{K_0}(q,p;E)=M(q,p;E) +\frac{8}{3\pi}
\ensuremath{\mathcal{P}}\int_0^\Lambda dq'\ \frac{q'^2
(\gamma+\sqrt{3q'^3/4-mE})}{q'^2-k^2} M(q',p;E) \tilde{K_0}(q,q';E),
\end{equation}
with $E$ and $k$ related as in Eq.~(\ref{eq:Eandk}).
The half-on-shell NLO K-matrix can then be expressed in terms of the
half-on-shell LO and NLO t-matrix and the on-shell LO t-matrix, LO
K-Matrix and NLO K-matrix
\begin{equation}
\tilde{K_1}(k,p;E)=\frac{\tilde{t_1}(k,p;E)}{1+i\frac{8\gamma k}{3}
\tilde{t_0}(k,k;E)} - i\frac{8\gamma k}{3} \tilde{t_0}(k,p;E)
\left[\tilde{K_1}(k,k;E) +\gamma \tilde{K_0}(k,k;E)\right],
\end{equation}
or vice versa
\begin{equation}
\label{eq:t1-K1-hs}
\tilde{t_1}(k,p;E) = \frac{\tilde{K_1}(k,p;E)}{1-i\frac{8\gamma k}{3}
\tilde{K_0}(k,k;E)}
+ \frac{ i\frac{8\gamma k}{3} \tilde{K_0}(k,p;E)
\left[\tilde{K_1}(k,k;E) +\gamma \tilde{K_0}(k,k;E)\right]}
{\left[1-i\frac{8\gamma k}{3} \tilde{K_0}(k,k;E)\right]^2}.
\end{equation}
At next-to-leading order, the half-on-shell K-matrix is then given by
the following principal-value integral:
\begin{eqnarray}
\label{eq:K1-hs}
\tilde{K_1}(k,p;E)&=&\frac{1}{\pi} \ensuremath{\mathcal{P}} \int_0^\Lambda dq'
q'^2\frac{\gamma+\sqrt{3q'^2/4-mE}}{-\gamma+\sqrt{3q'^2/4-mE}}\tilde{K_0}(k,
q';E) \tilde{K_0}(p,q';E)
\nn
&&+\frac{2 
\tilde{H_1}(\gamma,\Lambda)}{\Lambda^2}\left[1+\frac{2}{\pi}\ensuremath{\mathcal
{P}} \int_0^\Lambda
dq'\frac{q'^2}{-\gamma+\sqrt{3q'^2/4-mE}}\tilde{K_0}(k,q';E)\right]
\nn
&&\hspace{1.5cm}\times\left[1+\frac{2}{\pi} \ensuremath{\mathcal{P}}
\int_0^\Lambda
dq'\frac{q'^2}{-\gamma+\sqrt{3q'^2/4-mE}}\tilde{K_0}(p,q';E)\right].
\end{eqnarray}

\section{Three-body observables at NLO}
\label{sec:three-body-observ}
In this section we will discuss how different observables are
calculated at NLO in our perturbative approach. We will consider not
only obvious observables, such as phaseshifts and three-body binding
energies, but also observables typically
measured in experiments with ultracold atoms. We follow the strategy
outlined above and calculate all quantities as a series in powers of
$\gamma r_0$ and/or $k\, r_0$. The leading order in this series is
then the universal result, and the NLO pieces we will derive here
encode the first corrections ``beyond universality".

\subsection{Phaseshifts}
The amplitude $T(k)$ for atom-dimer scattering is related to the
atom-dimer S-wave phaseshift through
\begin{equation}
\label{eq:T-delta}
T(k)=\frac{3\pi}{m}\frac{1}{k\cot \delta(k)-ik}.
\end{equation}
The scattering amplitude $T(k)$ in Eq.~\eqref{eq:T-delta} can also be
expanded in powers of $\gamma r_0$
\begin{eqnarray}
\label{eq:T-cot}
T(k)&=&\frac{3\pi}{m}\frac{1}{k\cot\delta_0+r_0 [k\cot\delta]_1+\dots-ik}\nn
&=&\frac{3\pi}{m}\left[\frac{1}{k\cot\delta_0-ik}-r_0
\frac{[k\cot\delta]_1}{(k\cot\delta_0-ik)^2}+\dots\right]
\nn
&=&T_0(k)+T_1(k)+\dots,
\end{eqnarray}
where the dots refer to corrections beyond NLO. Here $k \cot \delta$ is expanded
as
\begin{equation}
k\cot\delta = k \cot\delta_0+ r_0 [k \cot\delta]_1+\dots,
\end{equation}
where the $[]_1$ indicates the part of $k \cot \delta$ that is the
coefficient of the order $r_0$ term in the expansion in powers of $r_0$.
At leading order we recover the familiar relation between phaseshifts
and K-matrix
\begin{equation}
\label{eq:K0-cot0}
k\cot\delta_0 = \frac{3}{8\gamma}\tilde{K_0}^{-1}(k,k;E),
\end{equation} 
but our expansion also leads to a relation for the NLO part, that stems from
Eqs.~\eqref{eq:T-cot} and
\eqref{eq:t1-K1-hs}:
\begin{eqnarray}
[k\cot\delta]_1
=-\frac{3}{8\gamma}
\tilde{K_0}^{-1}(k,k;E)\left(\gamma+\tilde{K_1}(k,k;E)/\tilde{K_0}(k,
k;E)\right).
\end{eqnarray}
As long as $k \cot \delta_0$ is not large this will yield a correction of
relative size $\gamma r_0$ to the leading part of $k\cot \delta$.

\subsection{Bound States}
The t-matrix is real for energies below the scattering threshold. The
NLO correction to the leading order t-matrix is then given by
\begin{eqnarray}
\label{eq:t1-rec}
\tilde{t_1}(q,p;E)
&=&\frac{1}{\pi}\int_0^\Lambda dq'
q'^2\frac{\gamma+\sqrt{3q'^2/4-mE}}{-\gamma+\sqrt{3q'^2/4-mE}}\tilde{t_0}(q,
q';E) \tilde{t_0}(p,q';E)\nn
&&+\frac{2
\tilde{H_1}(\gamma,\Lambda)}{\Lambda^2}\left[1+\frac{2}{\pi}\int_0^\Lambda
dq'\frac{q'^2}{-\gamma+\sqrt{3q'^2/4-mE}}\tilde{t_0}(q,q';E)\right]\nn
&&\hspace{2cm}\times\left[1+\frac{2}{\pi}\int_0^\Lambda
dq'\frac{q'^2}{-\gamma+\sqrt{3q'^2/4-mE}}\tilde{t_0}(p,q';E)\right]~,
\end{eqnarray}
where we have explicitly dropped the $i\epsilon$ prescription. 
The full t-matrix has a pole at $E=B$ if a three-body bound state
exists with this energy. At leading order this implies:
\begin{equation}
\label{eq:t0z0}
\tilde{t_0}(q,p;E)=\frac{\tilde{Z_0}(q,p)}{E-B_0}+\ensuremath{\mathcal{R}}_0(q,
p;E)
\end{equation}
where the function $\ensuremath{\mathcal{R}}_0$ is a regular part. The
residue $\tilde{Z_0}$ depends on the incoming and outgoing momenta $q$
and $p$, but not on the three-body energy $E$. Although, in general,
more than one bound state exists for the systems under consideration
here, the decomposition (\ref{eq:t0z0}) is the appropriate one if we
are focusing on the NLO shift for a particular bound state. The fact
that LO bound-state energies are separated by a factor as large as 515
allows us to employ the decomposition (\ref{eq:t0z0}) for these
purposes.

When considering the NLO correction, we must account for both the pole's
position (i.e. the three-body binding energy) and the residue being shifted by
an
amount proportional to $r_0$. Therefore we have
\begin{eqnarray}
\label{eq:tz}
\tilde{t_0}+r_0\tilde{t_1}&=&\frac{\tilde{Z_0}+
\tilde{Z_1}}{E-B_0-B_1}+\ensuremath{\mathcal{R}}_0 +
\ensuremath{\mathcal{R}}_1\nn
&=&\frac{\tilde{Z_0}}{E-B_0}+\frac{\tilde{Z_0}
B_1}{(E-B_0)^2}+\frac{\tilde{Z_1}}{E-B_0}+\ensuremath{\mathcal{R}}_0  +
\ensuremath{\mathcal{R}}_1,
\end{eqnarray}
where, as usual, the subscript 1 indicates the parts which are first
order in $r_0$. In particular, Eqs.~\eqref{eq:tz} and (\ref{eq:t0z0})
imply that the first-order part of $\tilde{t}$, $r_0 \tilde{t_1}$, has a
pole of order two at $E=B_0$:
\begin{equation}
\label{eq:t1z1}
r_0 \tilde{t_1}(q,p;E)=\frac{\tilde{Z_0}(q,p) B_1}{(E-B_0)^2}+
\frac{\tilde{Z_1}}{E-B_0} + \ensuremath{\mathcal{R}}_1(q,p;E).
\end{equation}
The residue of this double pole is then related to the shift in the
three-body binding energy that is linear in the effective range:
\begin{equation}
\label{eq:bz}
B_1=r_0\frac{\lim_{E\to B_0}(E-B_0)^2 \tilde{t_1}(q,p;E)}{\tilde{Z_0}(q,p)}.
\end{equation}
Equation~\eqref{eq:bz} seems to indicate that the incoming and outgoing
momenta, $q$ and $p$, affect the three-body binding energy shift
$B_1$. However, we would expect that the binding energy is independent of the
incoming and outgoing momenta. This apparent contradiction can be avoided if
$\tilde{Z_0}(q,p)$ is separable with respect to
$q$ and $p$. Therefore, the residue function is defined as
\begin{equation}
\label{eq:z0sep}
\tilde{Z_0}(q,p)=\Gamma(q)\Gamma(p).
\end{equation}
By substituting Eqs.~(\ref{eq:z0sep}) and (\ref{eq:t0z0}) into
Eq.~(\ref{eq:t0}) and taking the residue at $E=B_0$ we find
\begin{equation}
\label{eq:Gamma}
\Gamma(q)=\frac{2}{\pi}\int_0^\Lambda dq'\
M(q,q';E)\frac{q'^2}{-\gamma+\sqrt{3q'^2/4-mE}}\Gamma(q').
\end{equation}
The function $\Gamma(q)$ is thus a solution to a homogeneous integral
equation, and the overall normalization is not immediately
determined. We fix this normalization by the condition:
\begin{equation}
\Gamma^2(q)=\lim_{E\rightarrow B_0}(E-B_0)\tilde{t}_0(q,q;E).
\end{equation}

With $\Gamma(q)$ in hand, we can insert Eq.~\eqref{eq:z0sep} into
Eq.~\eqref{eq:t1-rec}, and multiply by $(E-B_0)^2$ and take the limit as
$E \rightarrow B_0$. This yields a result for $B_1$ that is
independent of $q$ and $p$:
\begin{eqnarray}
\label{eq:B1final}
B_1&=&\frac{r_0}{\pi}\int_0^\Lambda dq\
q^2\frac{\gamma+\sqrt{3q^2/4-mB_0}}{-\gamma+\sqrt{3q^2/4-mB_0}}\Gamma^2(q)\nn
&&+\frac{8 \tilde{H_1}(\Lambda) r_0}{(\pi\Lambda)^2}\left[\int_0^\Lambda dq\
\frac{q^2}{-\gamma+\sqrt{3q^2/4-mB_0}}\Gamma(q)\right]^2.
\end{eqnarray}
And indeed, in numerical calculations, Eqs.~\eqref{eq:bz} and
\eqref{eq:B1final} prove to be equivalent. $B_1$ can be obtained from
Eq.~\eqref{eq:bz} if desired, and the result found in that way is
independent of $q$ and $p$. 

\subsection{Three-Body Recombination}
Three-body recombination is a collision process in which three free
atoms combine into a dimer and an atom. The atoms can either recombine
into deeply bound two-body states ({\it deep dimers}) whose properties
cannot be described by the SREFT (but see Ref.~\cite{Braaten:2004rn}
for a discussion on how the effects of these deep states can be
included in the theory), or, provided the scattering length $a$ is
positive, into the two-body bound state which is explicitly included
in SREFT. The energy that is released when the two-atom bound state
forms is converted into kinetic energy and atom and diatom are
lost from the trap. The loss rate of atoms in a cold atomic gas due to
three-body recombination into the shallow dimer is determined by the
scattering amplitude for the reaction $A + A + A \rightarrow A+D$, as
we shall now show. The scattering-length dependence of the loss rate
therefore provides an experimental signature of Efimov physics in
trapped systems of ultracold atoms.

The atom loss rate is expressed as
\begin{equation} \frac{d n}{d
  t}=-3\frac{n^3}{3!}W_{fi},
\end{equation}
where $n$ is the number density of free atoms. 
(The factor of 3 arises due to the
loss of three atoms in each recombination event.)
According to Fermi's golden rule,
\begin{equation}
W_{fi}=2\pi|T(p_f)|^2 \frac{d \nu_f}{d E_f},
\end{equation} 
where $T$ is the amplitude for three-atom recombination: $A + A + A \rightarrow
A + D$, 
and the density of atom-dimer states $d\nu_f$ is
\begin{equation}
d\nu_f=\frac{d^3 p_f}{(2\pi)^3}.
\end{equation}
The kinetic energy at a momentum $p_f$ in the atom-dimer system is
\begin{equation}
E_f=\frac{{p_f}^2}{2m} + \frac{{p_f}^2}{4m}=\frac{3 {p_f}^2}{4m},
\end{equation}
and so the transition rate becomes
\begin{equation}
W_{fi}=\frac{2m}{3\pi}p_f|t_{\rm rec}(p_f)|^2.
\end{equation}

The recombination rate $\alpha$ is conventionally defined as 
\begin{equation}
\frac{\hbox{d}n}{\hbox{d}t}=-3\alpha n^3,
\end{equation}
and so
\begin{equation}
\label{eq:alpha}
\alpha=\frac{m}{9\pi}p_f|t_{\rm rec}(p_f)|^2. 
\end{equation}
At zero temperature three-body recombination takes place at the three-atom threshold
and the value of the relative momentum $p_f$ in the atom-dimer
system is therefore $2 \gamma/\sqrt{3}$.

\begin{figure}[tbpic:3reclo]
\centerline{\includegraphics[width=9cm,angle=0,clip=true]{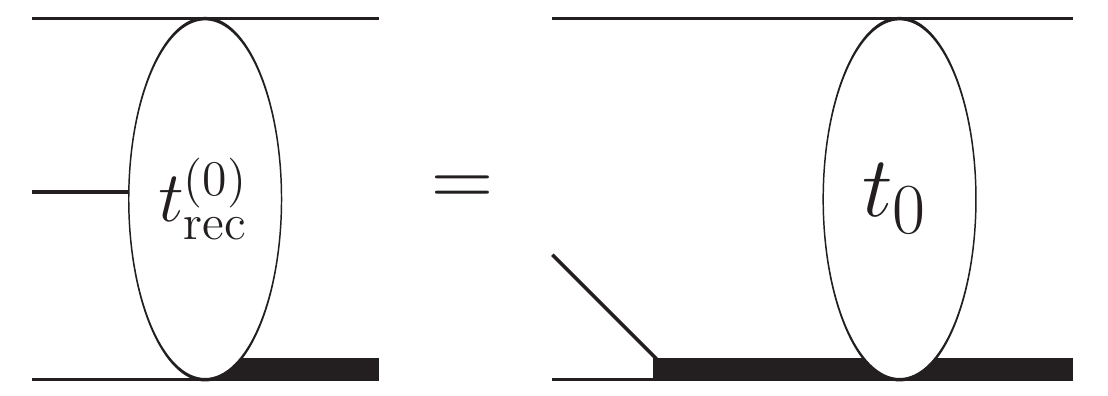}}
\caption{Three-body recombination at leading order}
\label{pic:3reclo}
\end{figure}

The $A + A + A \rightarrow A + D$ amplitude at leading order in
Fig.~\ref{pic:3reclo} is
related to the half-on-shell atom-dimer scattering t-matrix by
\begin{eqnarray}
t_{\rm rec}^{(0)}(p_f)&=&3\cdot(-i\sqrt{2}g)\cdot i\mathcal{D}^{(0)}(0,0)
\cdot\sqrt{Z_0}t_0(0,p_f;0)\nn
&=&\frac{48
\pi^{\frac{3}{2}}}{m\sqrt{\gamma}}\tilde{t_0}\left(0,\frac{2\gamma}{\sqrt{3}}
;0\right),
\end{eqnarray}
which, combined with Eq.~\eqref{eq:alpha}, determines the leading-order
recombination rate as:
\begin{eqnarray}
\label{eq:alpha0}
\alpha_0&=&\frac{512\pi^2}{\sqrt{3}m}\left|\tilde{t_0}\left(0,\frac{2\gamma}{
\sqrt{3}};0\right)\right|^2.
\end{eqnarray}

\begin{figure}[tbpic:3recnlo]
\centerline{\includegraphics[width=11cm,angle=0,clip=true]{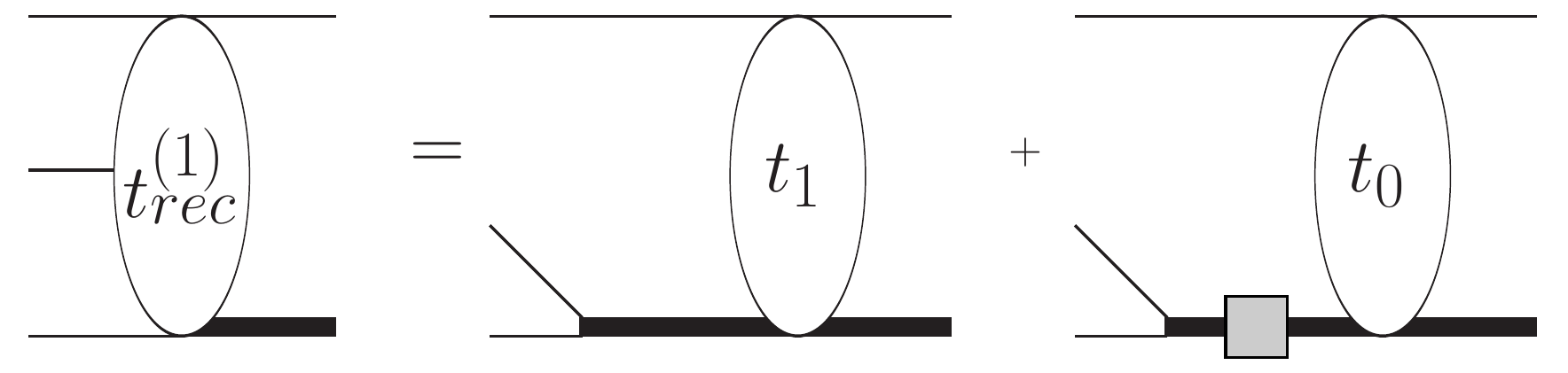}}
\caption{Three-body recombination at next-to-leading order}
\label{pic:3recnlo}
\end{figure}
Upon introducing the effective-range correction, we obtain the sum of the
leading-order
and next-to-leading-order $A + A + A \rightarrow A + D$ amplitude 
(Fig.\ref{pic:3recnlo}). This is expressed by
\begin{eqnarray}
t_{\rm rec}^{(0)}(p_f) + t_{\rm
rec}^{(1)}(p_f)&=&3\cdot(-i\sqrt{2}g)\sqrt{Z_0+Z_1}\cdot \left[
i\mathcal{D}^{(0)}(0,0)t_0 +i\mathcal{D}^{(1)}(0,0)t_0 +i\mathcal{D}^{(0)}(0,0)
t_1\right]\nn
&&\hspace{-2cm}=t_{\rm rec}^{(0)}+3\sqrt{2}g \left[\sqrt{Z_0}\
\mathcal{D}^{(1)}(0,0)t_0
+\frac{Z_1}{2\sqrt{Z_0}}\ \mathcal{D}^{(0)}(0,0)t_0 +\sqrt{Z_0}\
\mathcal{D}^{(0)}(0,0) t_1\right] + \ldots,
\nn
&=&t_{\rm rec}^{(0)}+\frac{48\pi^{3/2}}{m\gamma^{1/2}}\left[\gamma r_0
\tilde{t_0}+r_0\tilde{t_1}\right]\nn
\Rightarrow t_{\rm rec}^{(1)}(p_f)&=&\frac{48\pi^{3/2}}{m\gamma^{1/2}}
\left(\gamma r_0
  \tilde{t_0}(0,2\gamma/\sqrt{3};0)+r_0\tilde{t_1}(0,2\gamma/\sqrt{3};0)\right).
\end{eqnarray}
Therefore, the recombination rate is given at NLO by:
\begin{equation}
\label{eq:alpha1}
\alpha_0
+\alpha_1=\frac{512\pi^2}{\sqrt{3}m}\left|\left(\tilde{t_0}(0,\frac{2\gamma}{
\sqrt{3}};0)+\gamma r_0
\tilde{t_0}(0,\frac{2\gamma}{\sqrt{3}};0)+r_0\tilde{t_1}(0,\frac{2\gamma}{\sqrt{
3}};0)\right)\right|^2.
\end{equation}

\subsubsection{Recombination minimum}
\label{sec:recomb-minim}
At leading order the recombination rate $\alpha$ is related to
$\tilde{K}_0$ by
\begin{equation}
\alpha_0 = \frac{512\pi^2}{\sqrt{3}m} 
\frac{\tilde{K_0}^2(0,\frac{2\gamma}{\sqrt{3}};0)}{\left|1-i\frac{16\gamma^2
}{3\sqrt{3}}\tilde{K_0}(\frac{2\gamma}{\sqrt{3}},\frac{2\gamma}{\sqrt{3}}
;0)\right|^2}.
\end{equation}
Therefore, the recombination minimum is determined by the condition
$\tilde{K_0}(0,\frac{2\gamma_0}{\sqrt{3}};0)=0$, which leads to
$\tilde{t_0}(0,\frac{2\gamma_0}{\sqrt{3}};0)=0$.

Because $\alpha_0=0$ at the minimum, the NLO recombination rate in
Eq.~(\ref{eq:alpha1}) becomes $\ensuremath{\mathcal{O}}(r_0^2)$ in the
vicinity of the leading-order minimum. This means that in what follows
we cannot neglect terms $\sim r_0^2$, which complete the square and
guarantee that $\alpha>0$.

In order to calculate the NLO correction to the recombination minimum,
we evaluate $\alpha_{1}$ from Eq.~(\ref{eq:alpha1}) at
$\gamma=\gamma_0 +\Delta \gamma_0$, and expand it in powers of
$r_0$. First, we expand the scattering amplitude before we square it:
\begin{eqnarray}
\label{eq:t1-gamma0}
&&\left. \left[(1+\gamma r_0) \tilde{t_0}(0,\frac{2\gamma}{\sqrt{3}};0) +
r_0\tilde{t_1}(0,\frac{2\gamma}{\sqrt{3}};0)\right]\right|_{
\gamma=\gamma_0+\Delta \gamma}
\nn
&=&\left. \frac{\hbox{d}}{\hbox{d} \gamma}
\tilde{t_0}(0,\frac{2\gamma}{\sqrt{3}};0)\right|_{\gamma=\gamma_0} \Delta\gamma
+ r_0\tilde{t_1}(0,\frac{2\gamma_0}{\sqrt{3}};0)
+\ensuremath{\mathcal{O}}(r_0^2),
\end{eqnarray}
where $\Delta\gamma$ is linear in $r_0$, and we used the fact that
$\tilde{t_0}(0,\frac{2\gamma_0}{\sqrt{3}};0)=0$.

If we square the amplitude in Eq.~(\ref{eq:t1-gamma0}), we can
calculate $\alpha_1$ at $\gamma$ near $\gamma_0$ as
\begin{equation}
\alpha_1=\frac{512\pi^2}{\sqrt{3}m} \left|\left. \frac{\hbox{d}}{\hbox{d}
\gamma} \tilde{t_0}(0,\frac{2\gamma}{\sqrt{3}};0)\right|_{\gamma=\gamma_0}
\Delta\gamma + r_0\tilde{t_1}(0,\frac{2\gamma_0}{\sqrt{3}};0)\right|^2
+\ensuremath{\mathcal{O}}(r_0^3),
\end{equation}
where we used the fact that $\alpha_0=0$ at $\gamma=\gamma_0$. The
leading term here is quadratic in $r_0$ and purely determined by
up-to-NLO scattering amplitudes. Higher-order corrections to the
scattering amplitude only start to affect $\alpha$ at order
$r_0^3$. In other words, even though the NLO $\alpha$ near $\gamma_0$
is in the order of $r_0^2$, it is purely determined by LO and NLO
t-matrices.

From Eq.~(\ref{eq:t0-K0}) and Eq.~(\ref{eq:t1-K1-hs}), we derive
\begin{equation}
\left. \frac{\hbox{d}}{\hbox{d} \gamma}
\tilde{t_0}(0,\frac{2\gamma}{\sqrt{3}};0)\right|_{\gamma=\gamma_0} 
= \frac{1}{1-i\frac{16\gamma_0^2
}{3\sqrt{3}}\tilde{K_0}(\frac{2\gamma_0}{\sqrt{3}},\frac{2\gamma_0}{\sqrt{3}};0)
} \cdot \left.\frac{\hbox{d}}{\hbox{d} \gamma}
\tilde{K_0}(0,\frac{2\gamma}{\sqrt{3}};0) \right|_{\gamma=\gamma_0},
\end{equation}
and
\begin{equation}
\tilde{t_1}(0,\frac{2\gamma_0}{\sqrt{3}};0) =
\frac{\tilde{K_1}(0,\frac{2\gamma_0}{\sqrt{3}};0)}{1-i\frac{16\gamma_0^2
}{3\sqrt{3}}\tilde{K_0}(\frac{2\gamma_0}{\sqrt{3}},\frac{2\gamma_0}{\sqrt{3}};0)
},
\end{equation}
where we again applied $\tilde{K_0}(0,\frac{2\gamma_0}{\sqrt{3}};0)=0$.

Therefore, $\alpha_1$ near $\gamma_0$ is
\begin{equation}
\alpha_1=\frac{512\pi^2}{\sqrt{3}m} \frac{\left[
\left.\frac{\hbox{d}}{\hbox{d} \gamma} \tilde{K_0}(0,\frac{2\gamma}{\sqrt{3}};0)
\right|_{\gamma=\gamma_0}\Delta\gamma +
r_0\tilde{K_1}(0,\frac{2\gamma_0}{\sqrt{3}};0)\right]^2} {\left|
1-i\frac{16\gamma_0^2
}{3\sqrt{3}}\tilde{K_0}(\frac{2\gamma_0}{\sqrt{3}},\frac{2\gamma_0}{\sqrt{3}};0)
\right|^2}
\end{equation}

The next-to-leading-order recombination minimum $\alpha_1=0$ is thus
determined by
\begin{equation}
\left.\frac{\hbox{d}}{\hbox{d} \gamma}
\tilde{K_0}\left(0,\frac{2\gamma}{\sqrt{3}};0\right)
\right|_{\gamma=\gamma_0}\Delta\gamma_0 +
r_0\tilde{K_1}\left(0,\frac{2\gamma_0}{\sqrt{3}};0\right)=0,
\end{equation}
which leads to an NLO shift in the position of the recombination minimum of:
\begin{equation}
\label{eq:dgamma0}
\Delta \gamma_0= -r_0
\frac{\tilde{K_1}(0,\frac{2\gamma_0}{\sqrt{3}};0)}{\left.\frac{\hbox{d}}{\hbox{d
} \gamma} \tilde{K_0}(0,\frac{2\gamma}{\sqrt{3}};0) \right|_{\gamma=\gamma_0}}.
\end{equation}

\subsection{Atom-Dimer Resonance}
\label{sec:atom-diat-reson}

The atom-dimer scattering length diverges for positive scattering
lengths for which $B=-\gamma^2/m$. At this value of the scattering
length, a three-body state lies exactly at the atom-diatom
threshold. This feature shows up as resonant behavior in the
atom-dimer relaxation rate, a process in which shallow dimers are
transfered through collision processes into deep dimers. The
two-body binding momenta for which these resonances occur are denoted
(at LO) by $\gamma_*$. They thus obey the relation
\begin{equation}
\label{eq:gammastardefn}
B_0(\gamma_*)=-\frac{\gamma_*^2}{m}.
\end{equation}
Discrete scale invariance in the leading-order
bound-state spectrum implies that if $\gamma_*$ is a solution of
Eq.~(\ref{eq:gammastardefn}), then so are the quantities $e^{n
  \pi/s_0} \gamma_*$. The scale invariance is softly broken by $r_0/a$
corrections. Here we will calculate the NLO corrections to
these $\gamma_*$'s.

We will assume that the position of a particular resonance is shifted
to $\gamma_* + \Delta \gamma_*$ at NLO. The three-body binding energy
at the atom-dimer threshold must then obey up to terms of
relative order $\ensuremath{\mathcal{O}}(\gamma_*^2 r_0^2)$:
\begin{equation}
B_0(\gamma_*+\Delta \gamma_*) + B_1(\gamma_*+\Delta
\gamma_*)=-\frac{(\gamma_*+\Delta
\gamma_*)^2}{m}.
\label{eq:gammastardefnNLO}
\end{equation} 
We now expand both sides of (\ref{eq:gammastardefnNLO}) in powers of
$r_0$ and retain only terms up to $\ensuremath{\mathcal{O}}(r_0)$:
\begin{equation}
\label{eq:B0-ad}
B_0(\gamma_*)+\frac{d B_0(\gamma)}{d \gamma}|_{\gamma=\gamma_*}\Delta
\gamma_*+B_1(\gamma_*)=-\frac{\gamma_*^2}{m}-2\frac{\gamma_*\Delta \gamma_*}{m}.
\end{equation}
Using Eq.~(\ref{eq:gammastardefn}) we find a NLO correction to $\gamma_*$
\begin{equation}
\label{eq:dgamma*}
\Delta\gamma_*=-\frac{m B_1(\gamma_*)}{2\gamma_* + m \frac{d B_0(\gamma)}{d
\gamma}|_{\gamma=\gamma_*}}.
\end{equation}
We note that Eq.~(\ref{eq:B1final}) implies that $\Delta \gamma_*$ is linear in
$r_0$.

However, calculating $\Delta \gamma_*$ according to
Eq.~\eqref{eq:dgamma*} results in numerical difficulties, because
$B_1(\gamma_*)$ and the denominator are both zero to within the
numerical accuracy of our calculation.

The fact that the denominator should go to zero is clear from the
expression given for $B_0(\gamma)$ in \cite{Braaten:2004rn}:
\begin{eqnarray}
\label{eq:kg-H}
\kappa &=& -H \sin \xi\nn
\gamma &=& H \cos \xi\nn
H &=& \left(e^{-\pi/s_0}\right)^{n-n_*} \kappa_*\exp[\Delta(\xi)/(2s_0)],
\end{eqnarray}
where $\kappa = \sqrt{-m B_0}$. When $\gamma$ is near $\gamma_*$ the
function $\Delta(\xi)$ can be expanded in power of
$(-\pi/4-\xi)^{1/2}$. For completeness we give the expression of
Ref.~\cite{Braaten:2004rn}, with the coefficients that were determined
numerically there:
\begin{equation}
\label{eq:Dxi-gamma*}
\xi \in \left[ -\frac{3 \pi}{8}, -\frac{\pi}{4}\right]
:\Delta=6.04-9.63(-\frac{\pi}{4}-\xi)^{1/2}+3.10(-\frac{\pi}{4}-\xi),
\end{equation}
where $\xi=-\pi/4$ corresponds to the three-body bound state crossing
the atom-dimer threshold at $\gamma=\gamma_*$.  We calculate
$d\kappa/d\gamma$ from Eq.~(\ref{eq:kg-H}) and find that at $\xi
=-\pi/4$,
\begin{equation}
\left. \frac{d\kappa}{d\gamma} \right|_{\gamma=\gamma_*} 
=
\left. \frac{\frac{dH}{d\xi} - H}{\frac{dH}{d\xi} + H}\right|_{\xi=-\pi/4}
=
\left. \frac{\frac{d\Delta}{d\xi} - 2s_0}{\frac{d\Delta}{d\xi} +
2s_0}\right|_{\xi=-\pi/4} =1,
\end{equation}
since $\frac{d \Delta}{d \xi}\rightarrow \infty$ when $\xi\rightarrow-\pi/4$. $d
B_0/d \gamma$ at the atom-dimer threshold is therefore
\begin{equation}
\left.m \frac{d B_0}{d \gamma}\right|_{\gamma=\gamma_*} = -2\kappa
\left.\frac{d\kappa}{d\gamma}\right|_{\gamma=\gamma_*} = -2\gamma_*.
\end{equation}
This indicates that the denominator in Eq.~(\ref{eq:dgamma*}) indeed
goes to zero at the point of interest, which causes an accuracy
problem in numerically calculating $\Delta \gamma_*$ from the
bound-state side.

We calculate therefore instead $\Delta \gamma_*$ from the scattering
amplitude, i.e. evaluate the atom-dimer K-matrix as a function of
$\gamma$ along the threshold line $E=-\gamma^2$. The LO three-body
scattering length $a_3^{(0)}$ is related to $\tilde{K_0}$ at this energy by
\begin{equation}
\label{eq:a-K}
a_3^{(0)} = - \frac{8 \gamma}{3} \tilde{K}_0(0,0;-\gamma^2).
\end{equation}
Now $a_3^{(0)}\rightarrow\infty$ at $\gamma=\gamma_*$, and so the on-shell
$\tilde{K_0}$ has a pole of order one at $\gamma=\gamma_*$:
\begin{equation}
\label{eq:K0-pole}
\tilde{K}_0 (0,0;-\gamma^2) =
\frac{Z_0^{ad}(\gamma_*)}{\gamma-\gamma_*}+\ensuremath{\mathcal{R}}_0(\gamma).
\end{equation}
Similarly the NLO shift of the position of the atom-dimer resonance
$\Delta \gamma_*$, can be written as
\begin{eqnarray}
\label{eq:K0+K1}
\tilde{K}_0(0,0;-\gamma^2) +\gamma r_0\tilde{K}_0(0,0;-\gamma^2) +
r_0\tilde{K}_1(0,0;-\gamma^2)
&=&\frac{Z_0^{ad}+Z_1^{ad}}{\gamma-\gamma_*-\Delta
\gamma_*}+\mathcal{R}_0(\gamma)+\mathcal{R}_1(\gamma)\nn
&&\hspace{-5cm}=\frac{Z_0^{ad}}{\gamma-\gamma_*}+\frac{Z_1^{ad}}{\gamma-\gamma_*
}
+ \Delta \gamma_*
\frac{Z_0^{ad}}{(\gamma-\gamma_*)^2}+\ensuremath{\mathcal{R}}_0(\gamma)
+\ensuremath{\mathcal{R}}_1(\gamma)~.\nonumber\\
\end{eqnarray}
The shift $\Delta \gamma_*$ is therefore calculated as:
\begin{equation}
\label{eq:dgamma*-sc}
\Delta \gamma_* = r_0\frac{\lim_{\gamma\to \gamma_*}(\gamma-\gamma_*)^2
\tilde{K}_1(0,0;-\gamma^2)}{Z_0^{ad}(\gamma_*)}.
\end{equation}
Our numerical calculation shows that both the numerator and the
denominator in this form are finite, and so calculating $\Delta
\gamma_*$ by evaluating the atom-dimer K-matrix at different
$\gamma$'s along the threshold line is an accurate procedure.

However, we still have to show that Eqs.~(\ref{eq:dgamma*}) and
(\ref{eq:dgamma*-sc}) are equivalent for calculating $\Delta
\gamma_*$. The bound-state form of the leading-order atom-dimer
amplitude near a bound state of energy $B_0(\gamma)$
(Eq.~(\ref{eq:t0z0})) is:
\begin{eqnarray}
\tilde{t}_0(0,0;-\gamma^2)&=&\frac{m
\tilde{Z}_0(0,0)}{-\gamma^2-B_0(\gamma)}+\ensuremath{\mathcal{R}}_0
\nn
&=&-\frac{m \tilde{Z}_0(0,0)}{(\gamma-\gamma_*)\left[2\gamma_*+\left.\frac{d
B_0}{d \gamma}\right|_{\gamma=\gamma_*}\right]}+\ensuremath{\rm (regular)}.
\end{eqnarray}
This relates $\tilde{Z_0}$ in Eq.~(\ref{eq:t0z0}) to $Z_0^{ad}$ in
Eq.~(\ref{eq:K0-pole}) as
\begin{equation}
\label{eq:z0-bd-ad}
\left.\tilde{Z}_0(0,0)\right|_{\gamma=\gamma_*} = -\left(2\gamma_*+\left.\frac{d
B_0}{d \gamma}\right|_{\gamma=\gamma_*}\right)Z_0^{ad}(\gamma_*).
\end{equation}
Similarly we expand (\ref{eq:t1z1}) about $\gamma_*$ and so relate
$\tilde{Z}_0 B_1$ to the numerator in Eq.~(\ref{eq:dgamma*-sc}) as:
\begin{equation}
\label{eq:z1-bd-ad}
\left.\tilde{Z}_0 B_1\right|_{\gamma=\gamma_*} = \left(2\gamma_*+\left.\frac{d
B_0}{d \gamma}\right|_{\gamma=\gamma_*}\right)^2 
 r_0 \lim_{\gamma\to \gamma_*}(\gamma-\gamma_*)^2 \tilde{K}_1(0,0;-\gamma^2),
\end{equation}
which explains why $B_1 \rightarrow 0$ as $\gamma \rightarrow
\gamma_*$, and, moreover, shows that the coefficient of this zero is
precisely what is needed to render the expressions obtained for
$\Delta \gamma_*$ from the bound-state and scattering-state side
equivalent.

\subsection{Three-atom resonance}
\label{sec:three-atom-resonance}
When a state in the three-body bound-state spectrum crosses the
zero-energy threshold at negative scattering length, three free atoms
can form a zero-energy trimer state. This phenomena is called a
three-atom resonance, and results in a maximum in the three-atom
recombination rate. It occurs at a value of $\gamma$ denoted by
$\gamma_{-}$. (Since $\gamma_{-}<0$ this does not correspond to the binding
momentum of a dimer, but it is still the inverse of the atom-atom
scattering length where this feature occurs.) At leading order the condition for
this three-atom
resonance is $B_0(\gamma_{-})=0$. Furthermore, discrete scale invariance of the
leading-order bound-state spectrum then results in values of $\gamma_{-}$
being related by the universal scaling factor $e^{\pi/s_0}$. (Here and
below $\gamma_{-}$ denotes the leading-order position of the three-atom
resonance.)

The NLO correction to $\gamma_{-}$ is found at the NLO zero-energy
threshold $(B_0+B_1)(\gamma_{-} +\Delta \gamma_{-})=0$. We thus have: 
\begin{equation}
\label{eq:B0-3a}
B_0(\gamma_{-})+\left.\frac{d B_0(\gamma)}{d
\gamma}\right|_{\gamma=\gamma_{-}}\Delta
\gamma_{-}+B_1(\gamma_{-})=0.
\end{equation}
As $B_0(\gamma_{-})=0$ is given at LO, we find the NLO correction to
$\gamma_{-}$,
\begin{equation}
\label{eq:dgamma'}
\Delta\gamma_{-}=-\frac{B_1(\gamma_{-})}{\left.\frac{d B_0(\gamma)}{d
\gamma}\right|_{\gamma=\gamma_{-}}},
\end{equation}
and so $\Delta \gamma_{-}$ is linear in $r_0$. Our numerical studies show
that neither the numerator nor the denominator in this equation are
equal to zero at $\gamma=\gamma_{-}$. Our result is:
 \begin{equation}
 \left.\frac{d B_0(\gamma)}{d
\gamma}\right|_{\gamma=\gamma_{-}}=-0.984 \gamma_{-},
\end{equation}
where the coefficient is calculated with a numerical accuracy of about
$10^{-3}$.
This is in contradiction to the form
provided in
Ref.~\cite{Braaten:2004rn}, which predicts $\left.\frac{d B_0(\gamma)}{d
\gamma}\right|_{\gamma=\gamma_{-}}=-2 \gamma_{-}$. (See Appendix~\ref{ap-Gqn}.)

\section{The subleading three-body force at NLO}
\label{sec:subl-three-body}
We will show in this section explicitly that the NLO counterterm
contains a scattering-length-dependent piece that will require a
second experimental datum for renormalization if
scattering-length-dependent processes are considered. To do this we
reconsider Eq.~\eqref{eq:B1final}, the expression for the NLO shift to
the binding energy:
\begin{eqnarray}
\label{eq:B1final2}
B_1&=&\frac{r_0}{\pi}\int^\Lambda dq\
q^2\frac{\gamma+\sqrt{3q^2/4-mB_0}}{-\gamma+\sqrt{3q^2/4-mB_0}}\Gamma^2(q)\nn
&&+\frac{8 \tilde{H_1}(\Lambda) r_0}{(\pi\Lambda)^2}\left[\int^\Lambda dq\
\frac{q^2}{-\gamma+\sqrt{3q^2/4-mB_0}}\Gamma(q)\right]^2.
\label{eq:B1rep}
\end{eqnarray}
We will use the divergence structure of this observable to determine the
behavior of $\tilde{H_1}(\Lambda)$ as a function of $\Lambda$, up to corrections
$\sim 1/\Lambda$.
The divergence structure of any other observable computed to NLO will be
similar, and so it suffices to perform this calculation for $B_1$. In
particular, we will expand both the explicit integrals and the behavior of
$\tilde{H_1}(\Lambda)$, in powers of $\Lambda$, and demand that the
linear-in-$\Lambda$ and $\log(\Lambda)$ divergences cancel. 

In order to perform this analysis we need to know the large-momentum behavior of
each term in Eq.~(\ref{eq:B1rep}).
At large momenta $q$ the function $\Gamma(q)$ is
known in the form of an expansion in powers of $\gamma/q$ (see
Appendix of Ref.~\cite{Bedaque:2002yg} where we have corrected an error in the
result for $z_1$):
\begin{equation}
\Gamma(q) \propto \frac{z_0}{q}+\frac{\gamma z_1}{q^2}+\ldots~,
\label{eq:Gammaqas}
\end{equation}
with
\begin{eqnarray}
z_0 &=& \sin\left(s_0\ln\frac{q}{\bar{\Lambda}}\right)\\
z_1 &=& \frac{2}{\sqrt{3}}|C_{-1}|\sin\left(s_0\ln\frac{q}{\bar{\Lambda}}+\arg
C_{-1}\right)
\end{eqnarray}
where
\begin{equation}
C_{-1}=\frac{I(is_0-1)}{1-I(is_0-1)}
\end{equation}
and
\begin{equation}
I(s)=\frac{8\sin(\frac{\pi s}{6})}{\sqrt{3}s \cos(\frac{\pi s}{2})}~.
\end{equation}

Inserting the asymptotic form of $\Gamma(q)$, \eqref{eq:Gammaqas}, up to $\sim
1/q^2$, into
Eq.~\eqref{eq:B1final2}, and evaluating the first integral
shows that $H_1$ has to absorb both a linear divergence and a logarithmic
divergence proportional to $\gamma$. In order to cancel these cutoff
dependencies we will thus write $\tilde{H}_1$ as
\begin{equation}
\tilde{H_1}=\Lambda h_{10}(\Lambda) +\gamma
h_{11}(\Lambda).
\label{eq:H1}
\end{equation}

Analytic expressions for both $h_{10}(\Lambda)$ and $h_{11}(\Lambda)$
can then be obtained by inserting the expansions (\ref{eq:H1}) and
(\ref{eq:Gammaqas}) in Eq.~\eqref{eq:B1rep}, while also expanding all
explicit functions of $q$ in powers of $\gamma/q$. In this way we
find:
\begin{eqnarray}
\zeta &=&
\frac{1}{\pi}\int^\Lambda dq\ q^2
\left(1+\frac{4\gamma}{\sqrt{3}q}\right)\frac{1}{q^2}
\left(z_0^2+\frac{2\gamma}{q}z_0z_1\right) \nn
&&+\frac{8\Lambda}{\pi^2\Lambda^2}
\left(h_{10}+\frac{\gamma}{\Lambda}h_{11}\right)
\left[\frac{2}{\sqrt{3}}\int^\Lambda dq\ q
  \left(1+\frac{2\gamma}{\sqrt{3}q}\right)\frac{1}{q}
  \left(z_0+\frac{\gamma}{q}z_1\right)\right]^2 \nn \nn
&&=\frac{1}{\pi}\int^\Lambda dq\
\left(z_0^2+\frac{4\gamma}{\sqrt{3}q}z_0^2+\frac{2\gamma}{q}z_0z_1\right)
\nn &&+\frac{8}{\pi^2\Lambda}
\left(h_{10}+\frac{\gamma}{\Lambda}h_{11}\right) \frac{4}{3}
\left[\int^\Lambda dq\ \left(z_0+\frac{2\gamma}{\sqrt{3}q}z_0 +
    \frac{\gamma}{q}z_1\right)\right]^2 + \ldots,
    \label{eq:lengthy}
\end{eqnarray}
where $\zeta$ is finite, the dots represent finite parts of the integration,
and $\int^\Lambda$ defines an integral that is regulated in the ultraviolet by a
cutoff $\Lambda$ and whose infrared regularization (if any) is unspecified. 

In order to simplify the notation we denote
integrals that contain a product of the $z$ functions by a $\mathcal{W}$. 
The first and second indices indicate the $z$-functions
in the integrand, while the third index gives the power of q that resides in the
denominator of the integrand, so
\begin{equation}
\mathcal{W}_{lmn} \equiv \frac{1}{\pi}\int^\Lambda dq\ \frac{z_l z_m}{q^n}~.
\end{equation}
In the same spirit of notational convenience and compactness we define:
\begin{equation}
\mathcal{Z}_{mn} \equiv \frac{1}{\pi}\int^\Lambda dq\ \frac{z_m}{q^n}~.
\end{equation}
All integrals ${\cal W}_{lmn}$ and ${\cal Z}_{mn}$ can be evaluated
analytically, and this is done in Appendix~\ref{sec:relevant-integrals}.

The divergences linear in $\Lambda$ in Eq.~(\ref{eq:lengthy}) are then cancelled
by requiring:
\begin{equation}
\mathcal{W}_{000}(\Lambda) + \frac{32 h_{10}}{3\Lambda}
\mathcal{Z}_{00}^2(\Lambda)=0~.
\label{eq:WZ-h10}
\end{equation}
Meanwhile, the divergence which is logarithmic in the cutoff is canceled by
\begin{eqnarray}
\nonumber
&&\hspace{-2cm} \frac{2}{\sqrt{3}}\mathcal{W}_{001}(\Lambda)
+\mathcal{W}_{011}(\Lambda)
+\frac{16 h_{11}}{3\Lambda^2} \mathcal{Z}_{00}^2(\Lambda)\\
&&
+\frac{32 h_{10}}{3\Lambda}\mathcal{Z}_{00}(\Lambda)\left(\frac{2}{
\sqrt{3}}\mathcal{Z}_{01}(\Lambda)+\mathcal{Z}_{11}(\Lambda)\right)
=0~.
\label{eq:WZ-h11}
\end{eqnarray}

By using the results for these integrals which are given in Appendix
\ref{sec:relevant-integrals}\footnote{The result for, e.g.
$\mathcal{Z}_{00}(\Lambda)$, would seem to neglect the effect of the
infrared regularization, which could affect the answer for $h_{11}$. However,
the combination of the infrared-regularization dependence of
$\mathcal{Z}_{01}(\Lambda)$ and $\mathcal{Z}_{11}(\Lambda)$ is zero, and those
of $\mathcal{W}_{001}(\Lambda)$ and $\mathcal{W}_{011}(\Lambda)$ can be
absorbed into the numerically fitted parameter $\mu$, which is included in
$h_{11}(\Lambda)$.}
we derive the analytic forms for the NLO piece of the three-body force.
First,
\begin{equation}
\label{eq:h10}
h_{10}(\Lambda)=-\frac{3\pi (1+s_0^2)}{64
  \sqrt{1+4s_0^2}}\frac{\sqrt{1+4s_0^2} 
- \cos \left(2 s_0 \ln (\Lambda/\bar{\Lambda}) - \arctan 2s_0\right)} {\sin^2
\left( s_0 \ln (\Lambda/\bar{\Lambda}) - \arctan s_0\right)}~.
\end{equation}
Since $\bar{\Lambda}$ is determined by the LO renormalization condition
(e.g. $\bar{\Lambda}=13.1\kappa_*$ in the unitary limit), Eq.~\eqref{eq:h10} is
a prediction for $h_{10}(\Lambda)$. Meanwhile, Eq.~\eqref{eq:WZ-h11} gives
\begin{eqnarray}
\label{eq:h11}
h_{11}(\Lambda)&=&-\frac{\sqrt{3}\pi (1+s_0^2)}{16}\ 
\frac{(1+|C_{-1}|\cos(\arg C_{-1}))}{\sin^2 \left(
s_0 \ln (\Lambda/\bar{\Lambda})-\arctan s_0\right)} \ln (\Lambda/\mu)
\nn
&&+\frac{\sqrt{3} \pi (1+s_0^2)}{32 s_0}\ \frac{\sin\left(2s_0\ln
    (\Lambda/\bar{\Lambda})\right)
+|C_{-1}|\sin\left(2s_0\ln(\Lambda/\bar{\Lambda}) + \arg
C_{-1}\right)}{\sin^2 \left( s_0 \ln (\Lambda/\bar{\Lambda})-\arctan s_0\right)}
\nn
&&-\frac{\sqrt{3} \pi (1+s_0^2)^{3/2}}{16 s_0}\  
\frac{\cos\left(s_0\ln(\Lambda/\bar{\Lambda})\right)
+|C_{-1}| \cos\left(s_0\ln (\Lambda/\bar{\Lambda})+\arg
C_{-1}\right)}{\sin^3 \left( s_0 \ln (\Lambda/\bar{\Lambda})-\arctan
s_0\right)}\times
\nn
&&\hspace{0.5cm} \left[1
-\frac{1}{\sqrt{1+4s_0^2}}\cos\left(2s_0\ln(\Lambda/\bar{\Lambda}
)-\arctan (2s_0)\right)\right]
\label{eq:h11result}
\end{eqnarray}
The $\mu$ in Eq.~\eqref{eq:h11} subsumes information on the finite
part of $\ensuremath{\mathcal{O}}(\Lambda^0)$, and its value is
determined by the renormalization conditions at NLO. Numerically this piece is
of the same order as $\ln\Lambda$, as long as $\Lambda$ is not extremely large.

\begin{figure}[tbpic:h10]
\centerline{\includegraphics[width=15cm,angle=0,clip=true]{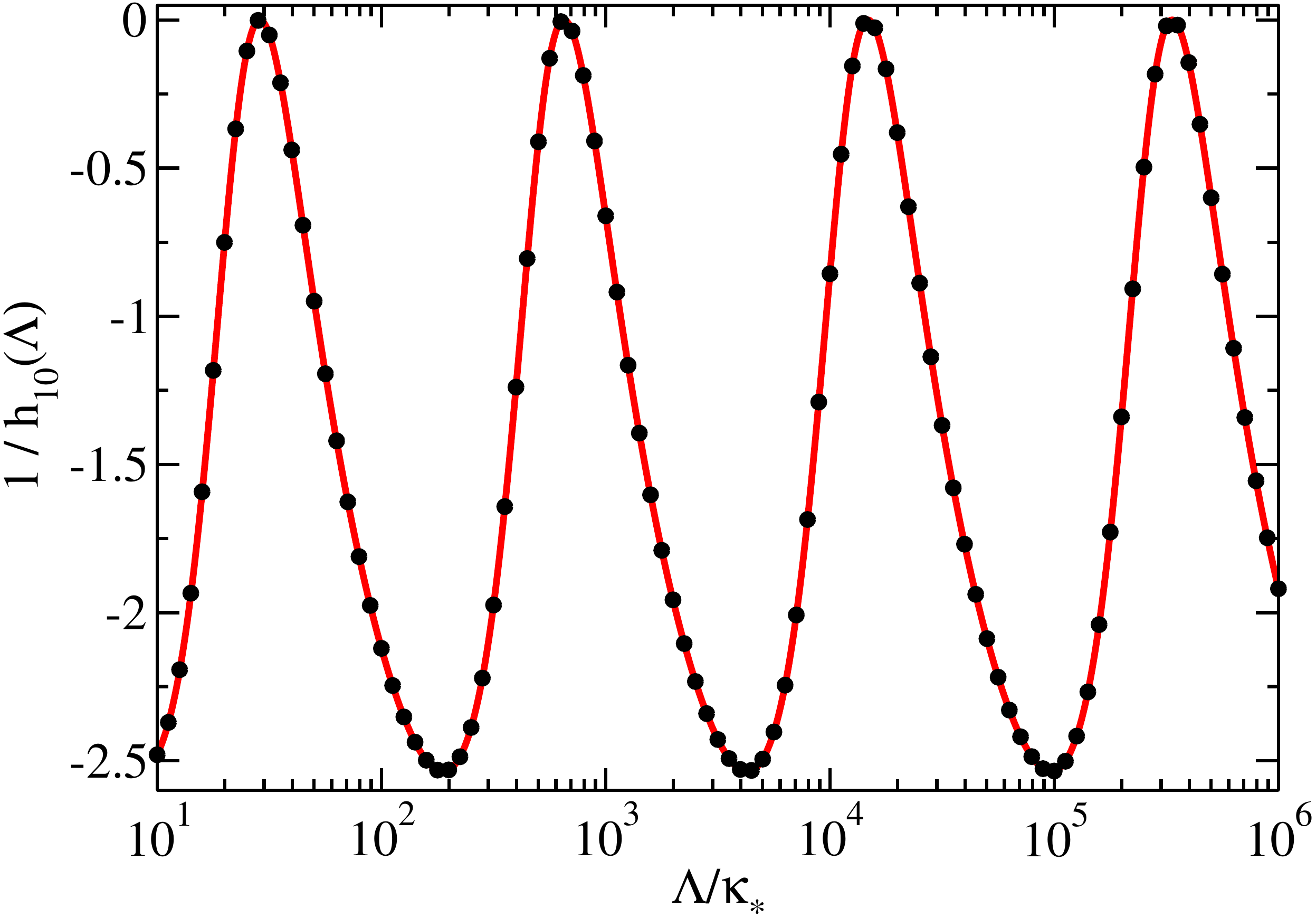}}
\caption{$h_{10}(\Lambda)$: 
The dots are numerical results for $h_{10}$ with the NLO calculation renormalized such
that the LO prediction of $\kappa_*=1$ ($\bar{\Lambda}=13.1$) is
maintained. The solid line (red) is the analytic function $h_{10}(\Lambda)$
given by Eq.~\eqref{eq:h10} with the same parameter $\bar{\Lambda}$. 
}
\label{pic:h10}
\end{figure}

\begin{figure}[tbpic:h11]
\centerline{\includegraphics[width=15cm,angle=0,clip=true]{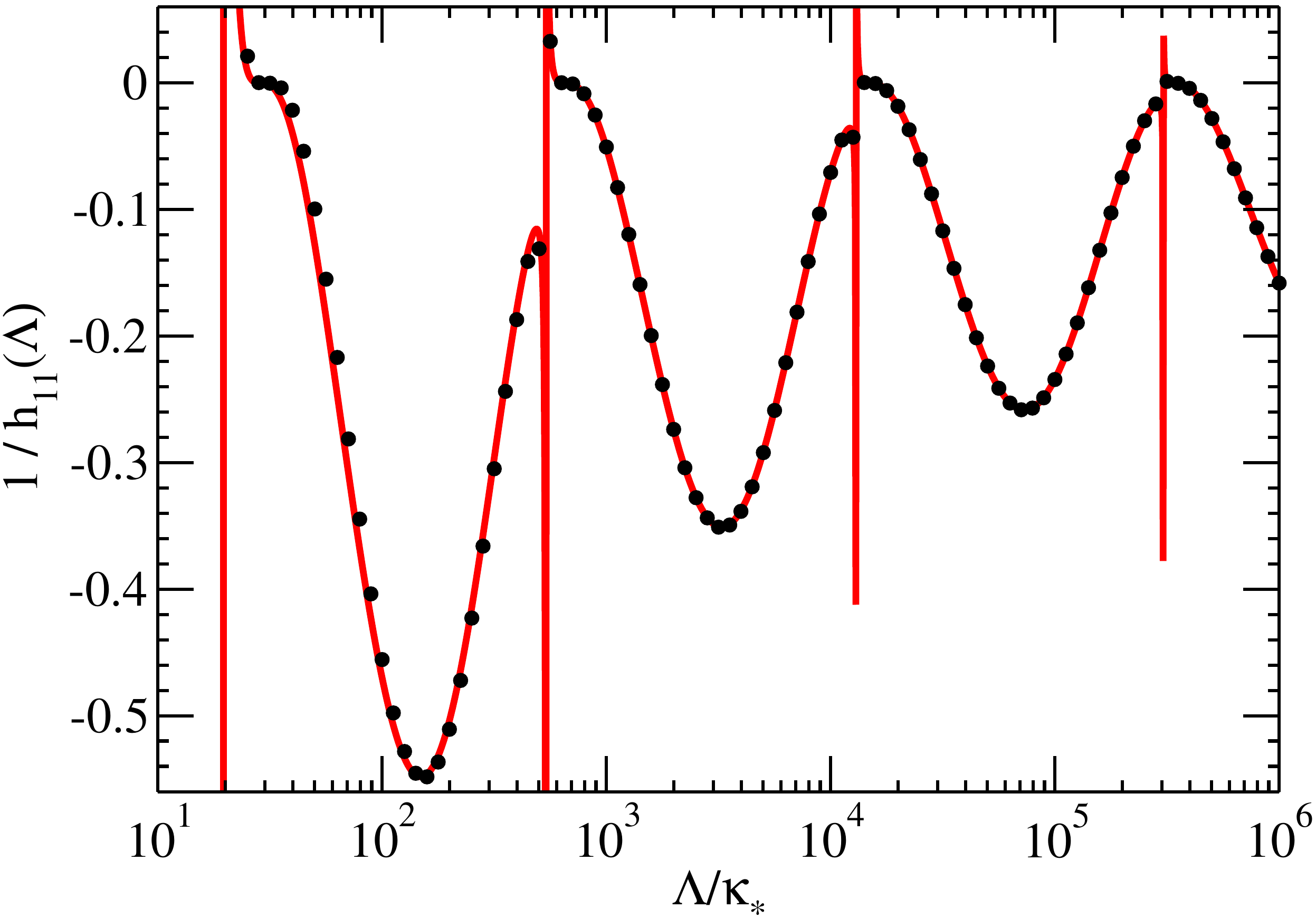}}
\caption{$h_{11}(\Lambda)$: The dots
are results of a numerical calculation of $h_{11}$ with the additional NLO counterterm
fitted to maintain the value of $\gamma_0 = \kappa_*/0.316$ predicted
at LO. The LO renormalization condition, $\kappa_*=1$
($\bar{\Lambda}=13.1$), is also kept at NLO. 
The solid line (red) is the analytic function $h_{11}(\Lambda)$
given by Eq.~\eqref{eq:h11}, with $\mu$ fitted to $0.99 \kappa_*$ 
in order to obtain good agreement with the numerical values.}
\label{pic:h11}
\end{figure}

We compare the analytical and numerical results for a particular
renormalization condition in Figs.~\ref{pic:h10} and
\ref{pic:h11}. 
Note that in order to do this comparison in Fig.~\ref{pic:h11} we have fitted the
value of $\mu$ to the 
numerical results. However, the coefficient of the
$\log(\Lambda/\mu)$ term in Eq.~(\ref{eq:h11result}) 
is still predictive, and is confirmed by comparison to the numerical
results.
The analytic and numerical results agree to within our numerical accuracy of $10^{-3}$ when
$\mu=0.990\kappa_*$.

The renormalization condition chosen here is $\kappa_*=1$ at LO and
NLO, and $\gamma_0=\kappa_*/0.316$ at NLO. This corresponds to a
recombination minimum that receives no range corrections at NLO.
The value of $\mu$ is affected by the choice of NLO renormalization conditions. 
We have checked that the numerical results for $h_{11}(\Lambda)$ obtained
with other renormalization conditions can be well described by choosing
alternative values of $\mu$.

In both figures we plot the inverse $1/h_{10}$ and $1/h_{11}$ of the
low-energy coefficients, and so the bumps that can be seen in
Fig.~\ref{pic:h11} are the result of zeros in Eq.~\eqref{eq:h11}.
The competition between the different terms in Eq.~(\ref{eq:h11result})
is clearly seen in the comparison. In particular, the double pole has a coefficient which grows
with $\ln \Lambda$, and this produces a turning point which is in increasing proximity to the position of the dominant
triple pole as $\Lambda$ grows.

In principle, the details of the regularization at NLO can affect the
discrepancy between analytical expressions and numerical results in the NLO
three-body force. In order to eliminate this effect, we first calculated the LO
amplitude, $t_0$, at a large cutoff, $\Lambda=10^{12}\kappa_*$. Then we
insert the resulted $t_0$ into NLO calculations integrated to a relatively
smaller cutoff, $\Lambda<10^6\kappa_*$, which suppresses the corrections from the
details of the regularization to $<10^{-6}$.

\section{Universal Relations at NLO}
\label{sec:nlo-rel}
In the limit $r_0/a=0$, the two-body scattering lengths at which key
experimental features occur, such as the atom-dimer
resonance at positive scattering length $a=a_*$, the three-atom
resonance at negative scattering length $a=a_{-}$, and the
recombination minimum at $a=a_0$, obey a set of universal
relations. Indeed, all three features can be related to the binding
momentum of the three-atom state in the unitary limit, $\kappa_*$
\cite{Braaten:2004rn}:
\begin{eqnarray}
\label{eq:uni1}
a_*&=&0.0708 \kappa_*^{-1}~,\\
a_{-}&=&-1.51 \kappa_*^{-1}~, \label{eq:uni2}\\
a_0&=&0.316 \kappa_*^{-1}~.  
\label{eq:uni3}
\end{eqnarray}

We can eliminate $\kappa_*$ in the equations above and then rewrite
them in terms of the inverse scattering length ($\gamma=1/a$ at LO)
\begin{equation}
\label{eq:uni-gamma*-gamma0}
\gamma_*=\theta_*\gamma_0~,\quad
\gamma_{-}=\theta_{-}\gamma_0~,\quad
\kappa_*=\theta_\infty\gamma_0~,
\end{equation}
where $\theta_*=4.47$, $\theta_{-}=-0.210$ and $\theta_\infty=0.316$.

Once the two-body effective range, $r_0$, is non-zero we can extend
the LO universal relations such as Eq.~\eqref{eq:uni-gamma*-gamma0} to
include its effects. However, as shown above, these
relations will now require two three-body parameters as input to
predict a third one. This means that, before predictions can be made,
two three-body observables have to be determined, either from
experiment or from the underlying microscopic interaction that governs
two- and three-body properties. To illustrate this, we assume that the
positions of the two features $\gamma_*$ and $\gamma_0$ are known. The
effects of the effective range {\it perturbation} can then be included
by adding a term linear in $r_0$ to the LO universal relation between
$\gamma_*$ and $\gamma_0$
\begin{eqnarray}
\label{eq:gam*NLO}
  \gamma_*=\theta_*\gamma_0+r_0 \,\mathcal{I}_* \gamma_0^2~.
\end{eqnarray}
The real and dimensionless number $\mathcal{I}_*$ is non-universal in the sense that it
depends on the renormalization condition, and thus on the type of
physical system under consideration. The universal relations between other
features are modified in the same way. For example
\begin{eqnarray}
\label{eq:gam-NLO}
  \Delta\gamma_{-}=r_0\,\mathcal{I}_{-} \gamma_0^2~,
\end{eqnarray}
where $\gamma_{-}$ can be predicted if the relation between $\mathcal{I}_*$ and
$\mathcal{I}_{-}$ is known.
Similarly, the shift from the LO value of $\kappa_*$ can be written as
\begin{eqnarray}
  \Delta\kappa_*=r_0\mathcal{I}_\infty \gamma_0^2~.
  \label{eq:Delkap*}
\end{eqnarray}
Note that in Eqs.~\eqref{eq:gam*NLO}--\eqref{eq:Delkap*} we have employed
$\gamma_0$ to make up the dimensions of the NLO correction. This is a matter of
convenience, since all dimensional quantities are related at LO by formulae such
as Eqs.~\eqref{eq:uni1}--\eqref{eq:uni3}. We also used the fact that the NLO
shifts in these three-body observables are strictly linear in $r_0$ in our
perturbative analysis.

While the parameters that we have introduced above are not universal,
the relations between them are. These are linear as a consequence of
first-order perturbation theory. We have obtained these linear relations
numerically and verified their stability for a wide range of possible
renormalization conditions. In the case of the renormalization scheme
discussed in the previous paragraph ($\gamma_0$ fixed at LO and NLO,
and $\gamma_*$ the additional input at NLO needed to fix $h_{11}$) we
found
\begin{eqnarray}
\label{eq:sruparams}
\nonumber
     \mathcal{I}_{-} &=&\xi_{-,*} + \eta_{-,*}\,\mathcal{I}_*~,\\
   \mathcal{I}_\infty &=&\xi_{\infty,*} +\eta_{\infty,*} \,\mathcal{I}_*~,
\end{eqnarray}
with
\begin{eqnarray}
  \label{eq:nlo-uni-params}
\nonumber
  \xi_{-,*}&=&0.256~,\\
\nonumber
\xi_{\infty,*}&=&-0.286~,\\
\nonumber
\eta_{-,*}&=&1.65\times10^{-2}~,\\
\eta_{\infty,*}&=&-2.04\times10^{-2}~.
\end{eqnarray}
Combining Eqs. \eqref{eq:gam*NLO}, \eqref{eq:gam-NLO},
\eqref{eq:Delkap*} and \eqref{eq:sruparams} with each other leads to
relations that contain only observable features and the parameters in
Eq.~\eqref{eq:nlo-uni-params}
\begin{subequations}
\begin{eqnarray}
 \gamma_{-} &=& \theta_{-}\gamma_0 + \xi_{-,*}\,r_0(a_{-})\,\gamma_0^2 + \eta_{-,*} \frac{r_0(a_{-})}{r_0(a_*)} 
\left(\gamma_* -\theta_*\gamma_0\right)~,\\
\kappa_* &=& \theta_\infty\gamma_0 + \xi_{\infty,*} \,r_0(a_{\infty})\,\gamma_0^2 + \eta_{\infty,*} \frac{r_0(a_{\infty})}{r_0(a_*)}
\left(\gamma_* -\theta_*\gamma_0\right)~.
\end{eqnarray}
\end{subequations}
It is important to note at this point, that different universal
relations can be generated depending on which three-body features are
assumed to be known.

We can now apply our results to recent experimental measurements of
recombination in ultracold gases. Lithium-7 atoms in the $|F=1\
m_F=0\rangle$ hyperfine state are one system for which the scattering
length and effective range are known as functions of the magnetic
field.  Gross {\it et al.}  \cite{Gross:2009} measured a number of
recombination features for this system. On the positive scattering
length side of the resonance, they were able to measure a
recombination minimum at $a_0\approx1160a_B$ and the position of an
atom-dimer resonance at $a_*\approx290a_B$. On the negative scattering
length side, they measured a recombination maximum
$a_{-}^{(-)}\approx-264a_B$. The superscript $^{(-)}$ indicates that this
$a_{-}$ represents the point where the next deeper trimer branch
relative to $a_0$ and $a_*$ crosses the three-atom threshold.
Ref.~\cite{Gross:2009} also contains the results of a calculation for
the magnetic field dependence of scattering length $a$ and effective
range $r_0$. We will employ these results to illustrate the use of the
relations presented above.

The values of the effective range at the scattering lengths where
experimental features occur are
\begin{eqnarray}
\nonumber
r_0(a_0) &=& -34.5 a_B\\
\nonumber
r_0(a_*) &=& -74.7 a_B\\
\nonumber
r_0\left(a_{-}^{(-)}\right) &=& 27.2 a_B, 
\end{eqnarray}
We now use the values of two of these observables to {\it predict} the third one
(whose measurement was already reported in Ref.~\cite{Gross:2009}).
In particular, we will predict $a_*$ after renormalizing to $a_0$ and
$a_{-}^{(-)}$.  In doing this we assume that the low-energy constants
$H_0$, $h_{10}$, and $h_{11}$ depend only weakly on the magnetic field
and are the same for all three experimental features we consider. We
start out again by relating the known features through the LO
universal relations and a shift linear in the effective range times a
non-universal number that contains the information about the NLO
renormalization condition
\begin{subequations}
\label{eq:gamma0-gamma*}
\begin{eqnarray}
\gamma_0 &=& -0.210\gamma_{-}^{(-)} + r_0(a_0)\cdot\mathcal{I}_0^{(-)}
{\gamma_{-}^{(-)}}^2~,\\
\gamma_* &=& -0.939\gamma_{-}^{(-)} + r_0(a_*)\cdot\mathcal{I}^{(-)}_*{\gamma_{-}^{(-)}}^2~,
\end{eqnarray}
\end{subequations}
We again can numerically extract the universal parameters that relate
the parameters $\mathcal{I}_0^{(-)}$ and $\mathcal{I}^{(-)}_*$ via
\begin{eqnarray}
  \mathcal{I}_*^{(-)}=-0.309+7.17\,\mathcal{I}_0^{(-)}~,
\end{eqnarray}
here we haven chosen to write out the numerical values of the
universal parameters directly to avoid notational clutter.
We can then combine these equations into one universal relation that
predicts the position of $a_*$.

The corresponding universal relation is given by
\begin{equation}
\label{eq:relation:final1}
\gamma_* = \theta_*^{(-)}\gamma^{(-)}_{-} -0.309\, r_0(a_*){\gamma^{(-)}_{-}}^2 + 7.17\, \frac{r_0(a_{*})}{r_0(a_{0})}
\left(\gamma_0 -\theta_0^{(-)}\gamma_{-}^{(-)}\right)
\end{equation}
where the parameters of the universal LO relations are given by
$\theta_*^{(-)} = -0.939$ and $\theta_0^{(-)}=-0.210$.

We can alternatively generate a relation that
gives the NLO result for $\gamma_*$ as the relative shift to the LO
universal equation between $\gamma_*$ and $\gamma_0$.
\begin{equation}
\label{eq:relation:final2}
\gamma_* = \theta_*\gamma_{0} -7.02\, r_0(a_*)\gamma_{0}^2 + 0.566 \frac{r_0(a_{*})}{r_0(a^{(-)}_{-})}
\left(\gamma^{(-)}_{-} -\theta_{-}^{(+)}\gamma_{0}\right)
\end{equation}
where $\theta_* = 4.47$, $\theta_{-}^{(+)} \equiv 1/\theta_0^{(-)}=-4.75$
and we have again immediately inserted the numerical values of the
NLO universal parameters. Equations \eqref{eq:relation:final1} and
\eqref{eq:relation:final2} will generally give different numerical
results for $\gamma_*$ which provides a lower bound on higher order
corrections.

Using the numerical values for the positions of three-body features
and the numerical values for the effective ranges, we obtain
\begin{eqnarray}
a_* &=& (271 - 108+\ldots) a_B~,\\
  a_* &=& (257 - 2+\ldots) a_B~,
\end{eqnarray}
from Eq.~\eqref{eq:relation:final1} and
Eq.~\eqref{eq:relation:final2}, respectively.

Our predictions of $a_*$ in both renormalization schemes are
consistent with our previous findings \cite{Ji:2010su}.
Combining both results and including a conservative estimate of higher
order corrections, we obtain therefore the same result as quoted in
Ref.~\cite{Ji:2010su}
\begin{eqnarray}
  a_*=(210\pm 44)a_B~.
\end{eqnarray}

\section{Conclusion}
\label{sec:conclusion}
In this work we have presented a perturbative calculation of
next-to-leading order (in $\ell/|a|$) corrections to universal
three-body physics. We have shown that an additional three-body
counterterm is required for renormalization if and only if
scattering-length-dependent quantities are considered. The inverse
scattering length therefore plays a similar role in the SREFT to that
of the pion mass in chiral perturbation theory. Counterterms
proportional to the inverse of the scattering length occur at orders
beyond leading, and these must be fitted by considering
scattering-length-dependent data.

The advantages of the perturbative analysis we have presented here are
twofold. First, it allowed us to derive analytic expressions for the
next-to-leading-order shifts of resonance positions in recombination
experiments. Second, it permitted an explicit treatment of the
renormalization of divergent integrals while keeping the LO
counterterm fixed. This produced analytic forms for the running of the
NLO pieces of the three-body force.

Our analysis is applicable to systems for which the scattering length
and effective range are known as a function of the magnetic field. If
data on three-body processes at different values of the two-body
scattering length exists then effective-range corrections to
recombination features can be treated in the manner described
above. Data from the Bar-Ilan and Rice
groups~\cite{Gross:2009,pollack:2009} on Lithium-7 recombination were
recently analyzed in this way~\cite{Ji:2010su}.

A calculation of the divergence structure of bosonic observables in SREFT
at next-to-next-to-leading
order ${\cal O}(\ell^2/a^2)$ is underway \cite{Ji:2011}.
\section*{Acknowledgments}
We thank Eric Braaten and Hans-Werner Hammer for useful discussions.
DRP and CJ are grateful for the hospitality of the HISKP at the
University of Bonn, and also thank the Institute for Nuclear Theory
at the University of Washington for its hospitality during the INT program
10-01 ``Simulations and Symmetries: Cold Atoms, QCD, and Few-nucleon
Systems''. Parts of this work were done in each of these venues.  This
work was supported by the US Department of Energy under contracts
DE-FG02-93ER40756 and DE-FG02-00ER41132, the Swedish Research
Council (LP) and by a Mercator Fellowship (DRP).
\appendix
\section{Relevant Integrals}
\label{sec:relevant-integrals}
The following functions are used in the derivation of the analytical expression
for the NLO three-body counterterms $h_{10}$ and $h_{11}$.
\begin{eqnarray}
  \label{eq:int-z01}
\hspace{-0.8cm}\mathcal{Z}_{00}&=&\frac{1}{\pi}\int^\Lambda dq\ z_0 =
\frac{1}{\pi}\int^\Lambda dq\ \sin\left(s_0\ln
\frac{q}{\bar{\Lambda}}\right)
=\frac{\Lambda}{\pi\sqrt{1+s_0^2}}\sin\left(s_0\ln
\frac{\Lambda}{\bar{\Lambda}}-\arctan s_0\right),
\end{eqnarray}
\begin{eqnarray}
\mathcal{Z}_{01}&=&\frac{1}{\pi}\int^\Lambda dq\ \frac{z_0}{q} =
\frac{1}{\pi} \int^\Lambda \frac{dq}{q}
\sin\left(s_0\ln\frac{q}{\bar{\Lambda}}\right)
= -\frac{1}{\pi s_0}\cos\left(s_0\ln\frac{\Lambda}{\bar{\Lambda}}\right),
\end{eqnarray}
\begin{eqnarray}
\mathcal{Z}_{11}&=&\frac{1}{\pi}\int^\Lambda dq\ \frac{z_1}{q} =
\frac{2|C_{-1}|}{\sqrt{3}\pi}\int^\Lambda
\frac{dq}{q} \sin\left(s_0\ln\frac{q}{\bar{\Lambda}}+\arg C_{-1}\right)\nn
&&= -\frac{2|C_{-1}|}{\sqrt{3}\pi
s_0}\cos\left(s_0\ln\frac{\Lambda}{\bar{\Lambda}}+\arg
C_{-1}\right)~.
\end{eqnarray}

Integrals that contain a product of the $z$ functions are denoted
by $\mathcal{W}$. The first and second index indicates the order of the two
$z$-functions 
in the integrand's numerator, and the third index gives power of q in its
denominator. I.e.
\begin{eqnarray}
\mathcal{W}_{000}&=&\frac{1}{\pi}\int^\Lambda dq\ z_0^2 = 
\frac{1}{2\pi}\int^\Lambda dq\
\left[1-\cos\left(2s_0\ln\frac{q}{\bar{\Lambda}}\right)\right]
\nn
&&= \frac{\Lambda}{2\pi}\left[1 -\frac{1}{\sqrt{1+4s_0^2}}
\cos\left(2s_0\ln\frac{\Lambda}{\bar{\Lambda}}-\arctan (2s_0)\right)\right],
\end{eqnarray}

\begin{eqnarray}
\mathcal{W}_{001}&=&\frac{1}{\pi}\int^\Lambda dq\ \frac{z_0^2}{q} =
\frac{1}{2\pi}\int^\Lambda
\frac{dq}{q} \left[1-\cos\left(2s_0\ln\frac{q}{\bar{\Lambda}}\right)\right]\nn
&&=
\frac{1}{2\pi}\left[\ln\Lambda-\frac{1}{2s_0}\sin\left(2s_0\ln\frac{\Lambda}{
\bar {
\Lambda}}\right)\right],
\end{eqnarray}

\begin{eqnarray}
\mathcal{W}_{011}&=&\frac{1}{\pi}\int^\Lambda dq\ \frac{z_0 z_1}{q} =
\frac{|C_{-1}|}{\sqrt{3}\pi}\int^\Lambda \frac{dq}{q} \left[\cos\left(\arg
C_{-1}\right)
- \cos\left(2s_0\ln\frac{q}{\bar{\Lambda}}+\arg C_{-1})\right)\right]\nn
&&= \frac{|C_{-1}|}{\sqrt{3}\pi}\left[\cos\left(\arg
C_{-1}\right)\ln\Lambda-\frac{1}{2s_0}\sin\left(2s_0\ln\frac{\Lambda}{\bar{
\Lambda}}+\arg C_{-1}\right)\right]~.
\end{eqnarray}

\section{Parameterization of the three-body spectrum on the
negative-scattering-length side}
\label{ap-Gqn}
We follow Ref.~\cite{Braaten:2004rn} and define $B_0$ on the negative
$\gamma$ side as
\begin{equation}
B_0(\gamma) = -\gamma^2 \ensuremath{\mathcal{G}}\left
(\frac{\gamma}{\gamma_{-}}\right),
\end{equation}
where the dimensionless function $\ensuremath{\mathcal{G}}(1)=0$ at
threshold.  By substituting this into Eq.~\eqref{eq:dgamma'} we have
\begin{equation}
\label{eq:dgamma'G}
\Delta \gamma_{-} = \frac{B_1(\gamma_{-})}{ \gamma_{-}
\ensuremath{\mathcal{G'}}(1)}~.
\end{equation}

The calculation of $\Delta \gamma_{-}$ is numerically accurate because
$\ensuremath{\mathcal{G'}}(1) \neq 0$. We find
$\ensuremath{\mathcal{G'}}(1)=0.98$. That
$\ensuremath{\mathcal{G'}}(1)\neq 0$ is supported by the empirical
fitting formula in \cite{Braaten:2004rn} near $\gamma=\gamma_{-}$, where
$\kappa$ and $\gamma$ obey Eq.~\eqref{eq:kg-H}, while the function
$\Delta(\xi)$ obeys a different expression from
Eq.~\eqref{eq:Dxi-gamma*}, because now we are near the three-atom
resonance, not the atom-dimer resonance, which is indicated by
$\xi=-\pi$:
\begin{eqnarray}
\label{eq:Dxi-gamma'}
\xi \in \left[ -\pi, -\frac{5\pi}{8}\right] &:&\Delta=-0.89+0.28z+0.25z^2
\nn
&&z=(\pi+\xi)^2\exp[-1/(\pi+\xi)^2],
\end{eqnarray}
$d B_0/d\gamma$ in this region is
\begin{eqnarray}
\frac{d B_0}{d\gamma}
=\frac{2H\sin\xi \left(\frac{1}{2s_0}\frac{d \Delta}{d \xi} \sin\xi +
\cos\xi\right)}{\frac{1}{2s_0}\frac{d \Delta}{d \xi}\cos\xi - \sin\xi},
\end{eqnarray} 
whose numerator and denominator both go to zero when $\xi \to -\pi$. After
applying $\rm{I'H\hat{o}}$pital's rule we calculate $dB_0/d \gamma$ at
$\gamma_{-}$:
\begin{equation}
\left.\frac{d
B_0}{d\gamma}\right|_{\gamma=\gamma_{-}}=\left.\frac{2H}{1-\frac{1}{2s_0}\frac{
d^2
\Delta}{d \xi^2}}\right|_{\xi=-\pi}.
\end{equation}
Equation~(\ref{eq:Dxi-gamma'}) indicates that $d^2 \Delta/d \xi^2=0$ at $\xi=-\pi$,
therefore
\begin{equation}
\label{eq:dBdg-gamma'}
\left.\frac{d B_0}{d\gamma}\right|_{\gamma=\gamma_{-}}=2H=-2\gamma_{-},
\end{equation}
and so
\begin{equation}
\label{eq:G1-gamma'}
\ensuremath{\mathcal{G'}}(1)=2.
\end{equation}
Thus, the function $\ensuremath{\mathcal{G'}}(1)$ is finite and nonzero from the
empirical fitting formula in \cite{Braaten:2004rn}. However, the value in
Eq.~\eqref{eq:G1-gamma'} is larger than the correct numerical value in Eq.~\eqref{eq:G1-gamma'}.


\end{document}